\documentstyle[12pt]{article}
\def\journal{\topmargin .3in	\oddsidemargin .5in
	\headheight 0pt	\headsep 0pt
	\textwidth 5.625in % 1.2 preprint size  %6.5in
	\textheight 8.25in % 1.2 preprint size 9in
	\marginparwidth 1.5in
	\parindent 2em
	\parskip .5ex plus .1ex		\jot = 1.5ex}
\journal

\catcode`\@=11
\def\marginnote#1{}
\newcount\hour
\newcount\minute
\newtoks\amorpm
\hour=\time\divide\hour by60
\minute=\time{\multiply\hour by60 \global\advance\minute by-\hour}
\edef\standardtime{{\ifnum\hour<12 \global\amorpm={am}%
	\else\global\amorpm={pm}\advance\hour by-12 \fi
	\ifnum\hour=0 \hour=12 \fi
	\number\hour:\ifnum\minute<10 0\fi\number\minute\the\amorpm}}
\edef\militarytime{\number\hour:\ifnum\minute<10 0\fi\number\minute}
\def\draftlabel#1{{\@bsphack\if@filesw {\let\thepage\relax
   \xdef\@gtempa{\write\@auxout{\string
      \newlabel{#1}{{\@currentlabel}{\thepage}}}}}\@gtempa
   \if@nobreak \ifvmode\nobreak\fi\fi\fi\@esphack}
	\gdef\@eqnlabel{#1}}
\def\@eqnlabel{}
\def\@vacuum{}
\def\draftmarginnote#1{\marginpar{\raggedright\scriptsize\tt#1}}
\def\draft{\oddsidemargin -.5truein
	\def\@oddfoot{\sl preliminary draft \hfil
	\rm\thepage\hfil\sl\today\quad\militarytime}
	\let\@evenfoot\@oddfoot	\overfullrule 3pt
	\let\label=\draftlabel
	\let\marginnote=\draftmarginnote
   \def\@eqnnum{(\theequation)\rlap{\kern\marginparsep\tt\@eqnlabel}%
\global\let\@eqnlabel\@vacuum}  }
\def\preprint{\twocolumn\sloppy\flushbottom\parindent 2em
	\leftmargini 2em\leftmarginv .5em\leftmarginvi .5em
	\oddsidemargin -.5in	\evensidemargin -.5in
	\columnsep .4in	\footheight 0pt
	\textwidth 10in	\topmargin  -.4in
	\headheight 12pt \topskip .4in
	\textheight 7.1in \footskip 0pt
	\def\@oddhead{\thepage\hfil\addtocounter{page}{1}\thepage}
	\let\@evenhead\@oddhead	\def\@oddfoot{}	\def\@evenfoot{} }
\def\numberbysection{\@addtoreset{equation}{section}
	\def\theequation{\thesection.\arabic{equation}}}
\def\underline#1{\relax\ifmmode\@@underline#1\else
	$\@@underline{\hbox{#1}}$\relax\fi}
\def\titlepage{\@restonecolfalse\if@twocolumn\@restonecoltrue\onecolumn
     \else \newpage \fi \thispagestyle{empty}\c@page\z@
	\def\thefootnote{\fnsymbol{footnote}} }
\def\endtitlepage{\if@restonecol\twocolumn \else \newpage \fi
	\def\thefootnote{\arabic{footnote}}
	\setcounter{footnote}{0}}  %\c@footnote\z@ }
\catcode`@=12
\relax
\def\figcap{\section*{Figure Captions\markboth
	{FIGURECAPTIONS}{FIGURECAPTIONS}}\list
	{Figure \arabic{enumi}:\hfill}{\settowidth\labelwidth{Figure 999:}
	\leftmargin\labelwidth
	\advance\leftmargin\labelsep\usecounter{enumi}}}
 \relax
\def\tablecap{\section*{Table Captions\markboth
	{TABLECAPTIONS}{TABLECAPTIONS}}\list
	{Table \arabic{enumi}:\hfill}{\settowidth\labelwidth{Table 999:}
	\leftmargin\labelwidth
	\advance\leftmargin\labelsep\usecounter{enumi}}}
 \relax
\def\reflist{\section*{References\markboth
	{REFLIST}{REFLIST}}\list
	{[\arabic{enumi}]\hfill}{\settowidth\labelwidth{[999]}
	\leftmargin\labelwidth
	\advance\leftmargin\labelsep\usecounter{enumi}}}
 \relax
\makeatletter
\newcounter{pubctr}
\def\publist{\@ifnextchar[{\@publist}{\@@publist}}
\def\@publist[#1]{\list
	{[\arabic{pubctr}]\hfill}{\settowidth\labelwidth{[999]}
	\leftmargin\labelwidth
	\advance\leftmargin\labelsep
	\@nmbrlisttrue\def\@listctr{pubctr}
	\setcounter{pubctr}{#1}\addtocounter{pubctr}{-1}}}
\def\@@publist{\list
	{[\arabic{pubctr}]\hfill}{\settowidth\labelwidth{[999]}
	\leftmargin\labelwidth
	\advance\leftmargin\labelsep
	\@nmbrlisttrue\def\@listctr{pubctr}}}
 \relax
\makeatother
\catcode`\@=11
\def\section{\@startsection {section}{1}{0pt}{-3.5ex plus -1ex minus
 -.2ex}{2.3ex plus .2ex}{\raggedright\large\bf}}
\catcode`\@=12
\newskip\humongous \humongous=0pt plus 1000pt minus 1000pt
\def\caja{\mathsurround=0pt}

\newif\ifdtup
\def\panorama{\global\dtuptrue \openup1\jot \caja
	\everycr{\noalign{\ifdtup \global\dtupfalse
	\vskip-\lineskiplimit \vskip\normallineskiplimit
	\else \penalty\interdisplaylinepenalty \fi}}}
\def\eqalignno#1{\panorama \tabskip=\humongous
	\halign to\displaywidth{\hfil$\displaystyle{##}$
	\tabskip=0pt&$\displaystyle{{}##}$\hfil
	\tabskip=\humongous&\llap{$##$}\tabskip=0pt
	\crcr#1\crcr}}
\def\oldreffmt#1{\rlap{[#1]} \hbox to 2\parindent{}}

\def\figfmt#1{\rlap{Figure {#1}} \hbox to 1in{}}

\def\abs#1{\left| #1\right|}

\def\beq{\begin{equation}}
\def\eeq{\end{equation}}
\def\ul{\underline}
\def\bea{\begin{eqnarray}}
\def\eea{\end{eqnarray}}
\catcode`\@=11
\def\eqnarray{\stepcounter{equation}\let\@currentlabel=\theequation
\global\@eqnswtrue
\global\@eqcnt\z@\tabskip\@centering\let\\=\@eqncr
\gdef\@@fix{}\def\eqno##1{\gdef\@@fix{##1}}%
$$\halign to \displaywidth\bgroup\@eqnsel\hskip\@centering
  $\displaystyle\tabskip\z@{##}$&\global\@eqcnt\@ne
  \hskip 2\arraycolsep \hfil${##}$\hfil
  &\global\@eqcnt\tw@ \hskip 2\arraycolsep $\displaystyle\tabskip\z@{##}$\hfil
   \tabskip\@centering&\llap{##}\tabskip\z@\cr}

\def\@@eqncr{\let\@tempa\relax
    \ifcase\@eqcnt \def\@tempa{& & &}\or \def\@tempa{& &}
      \else \def\@tempa{&}\fi
     \@tempa \if@eqnsw\@eqnnum\stepcounter{equation}\else\@@fix\gdef\@@fix{}\fi
     \global\@eqnswtrue\global\@eqcnt\z@\cr}

\catcode`\@=12
\font\tenbifull=cmmib10 % bold math italic
\font\tenbimed=cmmib10 scaled 800
\font\tenbismall=cmmib10 scaled 666
\relax

\def\thefootnote{\fnsymbol{footnote}}
\def\ref#1{$^{#1)}$}
\def\ul{\underline}
\def\newline{\hfil\break}
\def\versus{{\it versus}~}

\def\mtop{m_t}
\def\mb{m_b}
\def\mtau{m_\tau}
\def\mc{m_c}
\def\ms{m_s}
\def\mmu{m_\mu}

\def\me{m_e}
\def\Vcb{V_{cb}}
\def\Vub{V_{ub}}

\def\sinc{{\rm sin}\theta_C}

\def\utod{(m_u/m_d)}
\def\stod{(m_s/m_d)}

\def\tanb{{\rm tan}\beta}
\def\MGUT{M_{GUT}}

\def\althreemZ{\alpha_3(M_Z)}

\def\GeV{{\rm GeV}}
\def\MeV{{\rm MeV}}

\def\gapp{\mathrel{\raise.3ex\hbox{$>$}\mkern-14mu
              \lower0.6ex\hbox{$\sim$}}}

\def\lapp{\mathrel{\raise.3ex\hbox{$<$}\mkern-14mu
              \lower0.6ex\hbox{$\sim$}}}
\def\lsim{\lapp}
\def\hbar{{\mathchar'26\kern-.5em{\it h}}}

\begin{document}
\begin{titlepage}
\begin{center}
\today     \hfill    LBL-33531 \\
           \hfill    UCB-PTH-93/03 \\

\vskip .5in

{\large \bf A Systematic $SO(10)$ Operator Analysis For Fermion Masses.}
\footnote{This work was supported by the Director, Office of Energy
Research, Office of High Energy and Nuclear Physics, Division of High
Energy Physics of the U.S. Department of Energy under Contracts
DE-AC03-76SF00098 and DE-ER-01545-585, and by NSF grants PHY-92-19345 and
PHY-90-21139.}
%alternate footnote for faculty:
%\footnote{This work was supported in part by the Director, Office of
%Energy Research, Office of High Energy and Nuclear Physics, Division of
%High Energy Physics of the U.S. Department of Energy under Contract
%DE-AC03-76SF00098 and in part by the National Science Foundation under
%grant PHY90-21139.}

\vskip .5in
G. Anderson \footnote { Address after August, 1993:
Department of Physics, Massachusetts Inst. of Technology,
Cambridge, MA 02139}
and S. Raby \footnote{On leave, Theory Division, Los Alamos National Lab.,
Los Alamos, NM 87545.}\\
{\em Department of Physics, The Ohio State University\\
Columbus, OH 43210.}\\
\vskip .25in
S. Dimopoulos\\
{\em Department of Physics\\
Stanford university, Stanford, CA94305.}\\
\vskip .25in
L. J. Hall\\

{\em  Department of Physics, University of California\\
      and\\
      Theoretical Physics Group, Physics Division\\
      Lawrence Berkeley Laboratory, 1 Cyclotron Road\\
      Berkeley, California 94720}
\vskip .25in
G.D. Starkman\\
{\em Canadian Institute for Theoretical Astrophysics\\
University of Toronto\\
60 George St, Toronto, Ontario M5S 1A7, Canada.}\\

\end{center}
\newpage
\vskip .5in

\begin{abstract}
%insert abstract here
A new approach for deducing the theory of fermion masses at the scale of grand
unification is proposed. Combining SO(10) grand unification, family symmetries
and supersymmetry with a systematic operator analysis, the minimal set of
fermion mass operators consistent with low energy data is determined.
Exploiting the full power of SO(10) to relate up, down and charged lepton mass
matrices, we obtain predictions for 7 of the mass and mixing parameters.
The assumptions upon which the operator search and resulting predictions are
based are stressed, together with a discussion of how the predictions are
affected by a relaxation of some of the assumptions.
The
masses of the heaviest generation, $m_t,m_b$ and $m_\tau$, are generated from a
single renormalizable Yukawa interaction, while the lighter masses and the
mixing angles are generated by non-renormalizable operators of the grand
unified theory. The hierarchy of masses and mixing angles is thereby related to
the ratio of grand to Planck scales, $M_G / M_P$. An explicit realization of
the origin of such an economical pattern of operators is given in terms of a
set of spontaneously broken family symmetries. In the preferred models the top
quark is found to be heavy: $M_t = 180 \pm 15$ GeV, and $\tan \beta$ is
predicted to be very large.
Predictions are also given
for $m_s, m_s/m_d , m_u/m_d, V_{cb}, V_{ub}/V_{cb}$ and the amount of CP
violation. Stringent tests of these theories will be achieved by more precise
measurements of $M_t, V_{cb}, \alpha_s$ and $ V_{ub}/V_{cb}$ and by
measurements of CP violation in neutral B meson decays.

\end{abstract}
\end{titlepage}

\newpage
\renewcommand{\thepage}{\arabic{page}}
\setcounter{page}{1}
%THIS IS PAGE 1 (INSERT TEXT OF REPORT HERE)

\noindent{\bf  (1) INTRODUCTION}
\vskip 9pt

The Standard Model is unlikely to be a fundamental theory: it contains 19
parameters, 13 of which belong to the flavor sector of fermion masses and
mixing angles.
Two decades of attempts to incorporate the Standard Model into a  more
fundamental and economical theory have resulted in just {\em {one}}
highly significant quantitative success:
the calculation of sin$^2\theta_W$ in supersymmetric grand unified theories
[GUTs] \cite{GG} \cite{DRW}.
Although this is just one number, the quantitative agreement between theory
and experiment is so precise  that it constitutes an experimental hint
in favor of low energy supersymmetry.
Furthermore the prediction from supersymmetric GUTs has a much higher numerical
significance than the prediction from superstrings \cite{RB}.

The biggest obstacle to constructing a predictive theory is the flavor, or
fermion mass,
problem. Thirteen experimentally determined numbers are needed
to phenomenologically parameterize this sector of the theory.  But  to make
matters worse, with only 13 observables, the flavor sector is
a highly underdetermined system, since, typically, many more
parameters are needed to describe the Yukawa sector of some more
fundamental theory.  In order to make progress in this situation,  one
hopes that patterns in the masses and mixing angles can be explained by a
few fundamental parameters.  Tools which have been  used to tackle this tough
problem (and which minimize the number of fundamental parameters) include grand
unification and family symmetries. Grand Unification can, in principle,
relate the lepton, up and down mass matrices and hence reduce the
number of input masses by a factor $\sim 3$. Family symmetry  can enforce some
zero entries in the Yukawa matrices, again limiting the number of
fundamental parameters.

The consequences of the Georgi-Jarlskog ansatz\cite{GJ} for Yukawa
couplings in a supersymmetric theory were recently derived\cite{DHR}.
The ansatz included 7 parameters in the Yukawa
matrices plus one additional parameter entering the mass matrices, $\tan
\beta$, the ratio of the VEVs of the 2 Higgs present in any supersymmetric
theory.  With 8 parameters, there are 5 predictions for fermion masses
and mixing angles, and $\tan\beta$ is also predicted.
These predictions are encouraging: in
cases where they are already tested they agree with the data at the 90\%
confidence level or better \cite{DHR,ADHR}.

However, there are some shortcomings to this ansatz:

$\bullet$ While down quark and charged lepton mass matrices are related, there
are no relations between up and down quark masses,
even within the context of an SO(10) \cite{G} theory.

$\bullet$  The ansatz parameterizes the fermion mass hierarchy (eg. $m_{\tau}
\gg m_{\mu} \gg m_e$), but it provides no understanding of the origin of this
hierarchy.

$\bullet$  Although the ansatz can be obtained from a set of family symmetries,
it is nevertheless apparently adhoc. Perhaps there are other equally successful
ansa\"tze \cite{RRR}.

$\bullet$  There are 7  parameters in the Yukawa matrices at
M$_{GUT}$.  Can we find a theory with fewer parameters and thus more
predictions?

Our approach builds on an earlier attempt to construct very
predictive theories of fermion masses \cite{D83}, in which 4
input parameters gave all quark and lepton masses and mixings.
Although these earlier attempts were not completely successful,  they did not
incorporate CP-violation and the top quark mass was too small,  they did
suggest a paradigm for theories of fermion masses which, as we show
in this paper, are quite successful.

In this paper we make {\it a systematic operator analysis of the most
predictive flavor sectors possible in supersymmetric grand unified SO(10)
theories.}\footnote{We do not include neutrino masses in our analysis, which
would require the addition of further operators.}
The shortcomings mentioned above are addressed as follows:

$\bullet$ We exploit the full power of the SO(10) gauge symmetry
obtaining relations between the up and down Yukawa matrices.

$\bullet$ The family hierarchy is related to the ratio of grand to Planck mass
scales: $M_G/M_P$.

$\bullet$ Instead of obtaining an example of a single predictive
theory, we perform a
systematic search for the most predictive flavor sectors within the SO(10)
framework. We give general arguments as to the maximum number of predictions,
and the search reveals all such theories.

$\bullet$ The minimal flavor structure involves
just 4 SO(10) invariant operators. Since there is a single irremovable phase,
the Yukawa matrices depend on only 5 parameters.
Hence these theories have three
more predictions than the Georgi-Jarlskog ansatz, and two more predictions than
its $SO(10)$ analog\cite{GJ,DHR}.

An analysis of the predictions reveals
that some of these theories are already experimentally disfavored,
while the others can be distinguished by future experiments.

Why should nature be so kind as to choose the GUT flavor sector to be
maximally predictive? Such flavor sectors may be the only ones which can be
significantly probed experimentally, but can they be motivated theoretically?
It may be that the dominance of just four SO(10) invariant operators is a
result of a set of family symmetries and the breaking pattern of these
symmetries. This is an old idea \cite{FN,GJ,D83,RRR}. Unfortunately there is
considerable freedom in choosing the family symmetries and the breaking
pattern. Perhaps one day this will be understood in terms of a string
compactification. There is certainly no guarantee that the set of symmetries
resulting from a string theory will be very simple \cite{GKMR}.
We give an explicit realization of how family symmetries can lead to our
models, at least for the heavy two generations, in Appendix 1.
A more comprehensive development of such theories will be given elsewhere
\cite{BHR}.

In Section 2 we discuss some virtues of the gauge group SO(10) and some
features of the SO(10) interactions which generate quark and lepton
interactions.
In section 3 we summarize the assumptions which lead to our class
of predictive theories. We discuss the number of predictions in these theories
and show that there are just two possible textures: the ``22'' and ``23''
textures. In section 4 we give the
renormalization group program which must be performed to derive the predictions
for these theories, and discuss certain features common to both textures.
In section 5 an analytic analysis of the ``22''
texture is made, and nine models are discovered.
The seven predictions of
each of these models are derived using analytic approximations.
Numerical predictions for the nine ``22'' models
are given in section 6, showing that some
are already disfavored while others are successful.
Section 7 describes the result of a numerical search for
theories with the ``23'' texture; although this texture necessarily has a
certain degree of tuning, three models are found which agree well with data.
Conclusions are given in section 8.
Extensions of our ideas, and certain proofs, are left to appendices.
\vskip 9pt
\noindent{\bf (2) FEATURES OF $SO(10)$.}
\vskip 9pt

All our analysis is done within the context of a supersymmetric GUT,
which breaks at the GUT scale to the minimal supersymmetric standard model
(MSSM), thus preserving the good prediction for sin$^2\theta_W$
\cite{DRW}. The choice of GUT is constrained by our desire to relate up
and down quark, as well as the charged lepton, mass matrices. The smallest
grand unified symmetry that accomplishes this is $SO(10)$, which has
all the 16 chiral states of quarks and leptons (including the
right-handed neutrino) that comprise one family fitting neatly into the 16
dimensional spinor representation. In this sense $SO(10)$ is quite
unique since bigger groups containing one or more families in a single
representation typically contain a plethora of unwanted and unobserved
particles. The 3 16plets associated with the 3 known families will be denoted
by 16$_1$, 16$_2$ and 16$_3$, the last being the heaviest.

Another virtue of $SO(10)$ is that both Higgs that occur in
minimal supersymmetric
theories can fit into {\em{one}} 10-dimensional representation of $SO(10)$.
This implies that the $SO(10)$ invariant Yukawa interaction
$$
O_{33} =  A  16_3\ 10\ 16_3\eqno(2.1)
$$
can give mass to the $\tau $ lepton, top and bottom quarks of the 3rd family
in terms of just {\em{one}} coupling, $A$. This simple operator, first studied
by  Ananthanarayan, Lazarides and
Shafi\cite{ALS}, allows a prediction of the top quark mass from the b quark
mass without any reference to the lighter generations. An interesting feature
of this interaction is that the observed value of $m_b/m_\tau$, forces the
coupling A to be large, of order unity. In fact the top quark mass in this
scenario must be heavier than about 165 GeV \cite{HRS}. The heaviness of the
top quark is an immediate consequence of the large value for the third
generation Yukawa coupling; now the question to ask is rather why the b and
$\tau$ are so light. The large value for $m_t/m_b$ can only be obtained by
having a large ratio of electroweak VEVs $\tan \beta \equiv v_2/v_1 \approx
m_t/m_b$ \cite{OP}.
Such a large value of $\tan \beta$ requires a moderate fine tune in the MSSM
which we do not discuss in this paper \cite{HRRS}.
However, large $\tan \beta$ does have several
important phenomenological consequences, one of which is the
potentially very large weak-scale
radiative correction to the down-type quark masses.
At first sight these radiative corrections appear so large that
they will destroy the top mass prediction. As discussed in appendix 4,
this is typically not the case \cite{HRS}, and the effects of such corrections
will be ignored in the results presented in this paper.

In addition to the Higgs decouplet, 10, and 3 families, 16$_1$, 16$_2$, and
16$_3$,  more Higgs multiplets are  certainly necessary to break $SO(10)$ down
to $SU(3) \times SU(2) \times U(1)$.  These Higgs multiplets may enter the
fermion mass matrices in  operators with dimension( $\ge$ 4). The smallest such
representations are the 45 and the 16. The 16 can reduce the rank, breaking
$SO(10)$ down to $SU(5)$; it could also contribute to neutrino masses. The 45
can participate in several stages of $SO(10)$ breaking depending on the
direction in which it points. In concert, a 45 , 16 and $\bar{16}$ can break
$SO(10)$ to $SU(3) \times SU(2) \times U(1)$.

Since the 45 is the adjoint
representation of $SO(10)$, its VEV can point in any direction in the space
spanned by the 45 generators of $SO(10)$ as long as it leaves the group
$SU(3) \times SU(2) \times U(1)$ unbroken. This means that the
45 VEV lies in the 2
dimensional subspace of $U(1)'$ generators of $SO(10)$ that commute with $SU(3)
\times SU(2) \times U(1)$. There are 4 special directions in this subspace:

(1) $\langle 45\rangle = v_{10}\ e^{i\phi_1}\ T_1  \equiv 45_1 $

Here $v_{10}\ e^{i\phi_1}$ is the magnitude and phase of the VEV and $T_1
\equiv X$ is the $SO(10)$ generator that commutes with the $SU(5)$
Georgi-Glashow subgroup (the $X$ quantum numbers of family members are shown
in table 1). The reason why this direction is special is that it can break
$SO(10)$ down to $SU(5)\times U(1)_X$ at the scale $v_{10}$.  The linearly
independent direction is then given by

(2) $\langle 45\rangle = v_5 \ e^{i\phi_{24}} \ T_{24} \equiv 45_{Y}$

Here $T_{24} \equiv Y$ is the hypercharge generator and therefore it
 is not  $SU(5)$ invariant.  The scale at which the three couplings,
$\alpha_1, \alpha_2 $ and $\alpha_3$ unify is by definition the GUT scale,
$M_{G}$. In this paper we assume
$v_{10} \ge v_{5}$, resulting in a symmetry breaking pattern
$$SO(10){\stackrel{v_{10}}{\longrightarrow}} SU(5)
{\stackrel{v_5}{\longrightarrow}} SU(3) \times SU(2) \times U(1)_Y  $$
where  $v_5 \equiv M_{GUT}$.
%Another symmetry breaking pattern, which we will not make use of, occurs
%if $v_{5} > v_{10} \approx v_{16}$, giving $$
%SO(10) \stackrel{v_{5}}{\longrightarrow} SU(3) \times SU(2)_L \times
%U(1)_{Y} \times U(1)_{X} \stackrel{v_{10}}{\longrightarrow} SU(3)
%\times SU(2)_L  \times U(1)_Y.$$
%In this case, although  $v_5 = M_{G}$,  it does not represent a minimal
%3-2-1 unification scheme which is consistent with the LEP data. Finally, if
% $v_{5} \approx v_{16} > v_{10} $, then once again $v_5 = M_{G}$.  In this
%case, the $v_{10}$ VEV serves no useful role; we will not consider this
%possibility further.

There is another special direction in this two dimensional subspace,

(3) $ \langle 45\rangle = v_5 \ e^{i\phi_{B-L}} \ T_{B-L} \equiv 45_{B-L}$

Here $B-L$ is simply ordinary baryon minus lepton number.  Although it is not
linearly independent of the previous two directions,  it may nevertheless
play a significant dynamical role. It has been suggested before that this VEV
might naturally induce the necessary doublet-triplet splitting in the Higgs
sector \cite{CHH}.  Once again, since this VEV breaks $SU(5)$,  we shall assume
that the magnitude of it's VEV is given by $v_5 = M_{G}$.  Finally, the
linearly independent direction is

(4) $\langle 45\rangle = v_5\ e^{i\phi_{3R}}\ T_{3R} \equiv 45_{T_{3R}}$.

$T_{3R}$ is the 3rd component of the right handed isospin group.
It also breaks $SU(5)$ and so we have again assumed that it has a VEV of
magnitude $v_5$.

In any complete $SO(10)$ model, additional Higgs in 54 or 210 dimensional
representations may be necessary to force the desired breaking pattern.  We
shall not consider these states in our analysis.
As discussed,  we have identified $v_5 = M_{G}$.  As for the scale,
$v_{10}$, we will assume that it can take values anywhere between
the Planck mass $M_P$ and $M_{G}$.

The effective theory near the scale $M_{G}$ will then consist of renormalizable
terms plus higher dimension  operators suppressed by powers of either
$v_5 / M_P,  v_{10} / M_P$ or $v_{5}/ v_{10}$\footnote{These higher dimension
operators may, in principle, be obtained by integrating out heavy states at
the scales $v_{10}$ or $M_P$.  See, appendix 1 for more details of such an
interpretation.}. In particular, there are higher dimension operators of the
form $$ O_{ij} \equiv 16_i\, {45_1\over M_1}. . .  {45_k\over M_k}\, 10\,
{45_{k+1}\over M_{k+1}} . . .
{45_\ell\over M_\ell}\, 16_j\eqno(2.2)
$$
where some of the $M_n$ in the denominator can be $\sim M_P$ and others
$\sim v_{10}$, and $i,j = 1,2,3$. In writing these operators we understand the
SO(10) group invariant to be formed as follows: use the SO(10) gamma matrices
to write the 45 and 10 as $16 \times 16$ matrices and then compute the
invariant by a succession of matrix multiplications. The operators so formed
are those which result from integrating out heavy 16 and $\bar{16}$, as shown
by an example in appendix 1. Other group contractions are possible, for example
by integrating out a heavy 144, but we do not include these other operators in
this paper. We seek the predictive flavor sectors which could result from only
the simplest GUTs.

When the 45s get VEVs (in any of the 4 preferred directions
$45_1, 45_Y, 45_{B-L}$, $45_{T_{3R}}$)
these operators contribute to the $ij$-th element of the
mass matrices of charged leptons, up and down quarks.
These contributions are very simply related to each other: every time a 45
VEV couples to a quark or lepton it just counts its $X,
Y, (B-L)$ or $T_{3R}$ charge (shown in table 1) depending on whether the
VEV points in the $45_1, 45_Y, 45_{(B-L)} $ or $45_{T_{3R}}$ direction.
Thus, armed with this table, and for any choice of the VEVs, we can easily
compute the contribution of the operator (2.2) to lepton or up and down quark
mass matrices.

\begin{center}
\begin{large}
\indent {\bf Table 1: Quantum Numbers}
\vskip 20pt
\begin{tabular}{|c|c|c|c|c|}
\hline
&$X$ &$ Y$&$B-L$&$T_{3R}$\cr
\hline
  $q$ &1 &1 & 1 &0\cr
$u^c$  &1 &-4&-1&1\cr
  $d^c$&-3 &2&-1&-1\cr
$\ell$&-3&-3&-3  &0\cr
$e^c$  &1&6 &3  &-1\cr
  $\nu^c$&5 &0&3  & 1\cr
\hline
\end{tabular}
\end{large}
\end{center}

In these models, the fermion mass hierarchy
originates from a hierarchy of mass scales: $ v_5 =
M_G  \le v_{10} \le M_P$.

\vskip 9pt
{\bf (3a) THE ASSUMPTIONS}
\vskip 9pt

Our objective is to construct the most economical and predictive class of
theories of quark and lepton masses. This clearly involves making a set of
assumptions, which we now summarize:

(1) We require a supersymmetric GUT to preserve the successful prediction for
sin$^2\theta_W$  , as discussed earlier, and $SO(10)$ is taken as the gauge
group, since it is the smallest group relating leptons, up and down quarks
with no superfluous particles.

(2) Beneath the GUT scale the effective theory is taken to be the MSSM.

(3)The 2 Higgs doublets that occur in the MSSM both belong
to a unique 10 of $SO(10)$.
This leads to an economy of parameters: for example the single Yukawa
coupling $A$ of equation (2.1) is responsible for the masses of three
fermions of the third generation: $t, b$ and $\tau$.

(4) All dimensionless couplings of the GUT should be of order unity.
This implies that
only the 3rd family, 16$_3$, gets a mass via a renormalizable operator
having dimension
$d\leq 4$. Given the previous assumption, this operator is necessarily that
of eq. (2.1).

(5) All lighter families get their mass via the higher dimension operators
$O_{ij}$ of eq. (2.2).
Recall that the $M_n$s in the denominators emerge from physics beyond
$M_{G}$ (for more details on this point, see Appendix 1).
Some of the $M_n$s are $\sim M_P$ and others are proportional to
$\langle 45_1\rangle$ and therefore $\sim v_{10}$.  The ratio of scales
 will provide a partial understanding of the fermion mass hierarchy.  These
ratios are bounded, since the lower scale $M_G \sim 10^{16} GeV$ and $M_P
\sim 10^{18} GeV$.

(6) Each of the $\langle 45_n\rangle$s (occurring in the numerator of the
operators $O_{ij}$ of eq. 2.2) can only point in one of
the 4 directions $45_1,
45_Y, 45_{(B-L)}$ and $ 45_{T_{3R}}$ introduced in section 2.

This hypothesis is often satisfied in specific $SO(10)$ models.
The reason is that, as mentioned in section 2, these VEVs accomplish a very
specific step on the breaking chain of $SO(10)$ down to $SU(3) \times
SU(2)\times U(1)$. In some cases this hypothesis is a corollary of the more
general Extended Survival Hypothesis \cite{ESH}.

(7) The charged lepton masses and the quark masses and mixings are assumed to
be described, to a certain level of accuracy, by the smallest number of SO(10)
invariant operators possible. Such a picture may emerge from a pattern of
spontaneously broken family symmetries.

(8) The parameters of the MSSM are taken such that the one-loop weak-scale
radiative corrections to the masses of the down-type quarks and charged leptons
can be neglected. We have argued in appendix 4 why this is expected to be true
when $\tan \beta$ is large.

\vskip 9pt
{\bf (3b) THE OPERATOR SEARCH}
\vskip 9pt

This is a long list of assumptions: without a fundamental theory of fermion
masses some such list is always necessary. We believe that each assumption is
reasonable and are encouraged by the success of the scheme. Some of the
assumptions, for example (3) and (7), are strong, but we should stress that not
all assumptions are needed for each prediction. In fact, for any one of our
predictions a weaker set of assumptions can be formulated. However, we believe
that we have given the minimal set of assumptions for all seven flavor
predictions to result from the same theory.

Using the above assumptions we can now begin to construct theories of
fermion masses. We first show that the minimal texture of this type
includes just four effective operators.

In the absence of the non-renormalizable operators $O_{ij}$ of eq. (2.2) it is
clear that only the 3rd generation $\tau, t$ and $b$ fermions will acquire
mass via the operator of eq. (2.1).
The physical quantities of interest in this sector are 4: $m_\tau, m_t$,
and $m_b$ and $\tan\beta$.
These are given in terms of the Yukawa coupling $A$ and $\tan\beta$.
Following ALS \cite{ALS},
we  use the experimental values for $m_b/m_\tau$ and $m_\tau$ to
fix $A$ at $M_{G}$.
We then predict $m_t$ and $\tan\beta$.
The results are shown in figure 1 as a function of $m_b$ for various values
of $\alpha_s(M_Z)$.
It is important to note that for $\alpha_s$ in the range  .11 $\leq
\alpha_s \leq .13$, the top mass comes out heavier than $\sim$ 160 GeV.
If the top is lighter than 160 GeV, then at least one of assumptions (3) and
(8) must be incorrect.

Now we come to the two lighter families.
It is clear that there must be
a minimum of 2 operators of the form of equation (2.2) in order for the mass
matrices to have a non-zero determinant, which is necessary to ensure that the
electron is not massless. If there were only two such operators, they would
necessarily be $O_{23}$ and $O_{12}$ which give rise to $V_{cb}$ and $V_{us}$
respectively.  Note that we cannot replace  $O_{12}$ with
$O_{13}$, since the resulting $V_{us}$ would be too small.  These
operators $O_{23}, O_{12}$ together with the Yukawa
coupling of eq. (2.1) are {\em{not}} enough: in appendix 2 we show that if
$O_{23}$ is of dimension 5 or 6, agreement with the observed masses and mixings
of the heavy two families cannot be obtained. It is shown that an additional
operator must contribute either to the 22 entry or to
the 23 entry. A corollary of this theorem is that our
theories have Kobayashi-Maskawa type CP-violation.
This is because the phases of the Yukawa coupling, A, and the 3 effective
Yukawa couplings, resulting from the higher dimension operators
cannot all be removed by redefining the
phases of $16_1, 16_2$ and $16_3$.    We thus conclude that the minimal
texture includes 4 effective operators,  which result in 6 arbitrary
parameters in the fermion mass matrices (4 magnitudes and one phase in the
Yukawa matrices and $\tan \beta$).  We will thus have 7 flavor
predictions.  We can
have the two possible textures defined by $$ O_{33} + O_{23} + O_{22}
+O_{12}   \eqno(3.1) $$ or
$$ O_{33} + O_{23}^a + O_{23}^b + O_{12}. \eqno(3.2) $$

\vskip 9pt
\noindent{\bf (4) RENORMALIZATION GROUP, $O_{33}$ AND $O_{12}$}
\vskip 9pt
In this section we consider the
RG evolution of the Yukawa eigenvalues and mixing angles from the GUT scale
to low energies.
We make use of well-known simple 1-loop formulas to scale from grand to weak
scales; $M_G $ to $M_S$ \cite{GLT,POK}, which include the effects of large $t,
b$ and tau Yukawa couplings. These results will be used in the next section. In
the numerical analyses of sections 6 and 7 a two loop RG analysis is
performed. In this section we also discuss $O_{33}$ and $O_{12}$ which are
unique and therefore common to both textures.

The third generation Yukawas at the weak scale are
$$
\lambda_t = \lambda_{t_G}\mathop{\Pi}_{a} \zeta^u_a \: e^{-6I_t - I_b}
 \eqno(4.1a)
$$
$$
\lambda_b =\lambda_{b_G} \mathop{\Pi}_{a} \zeta^d_a
\: e^{-I_t - 6I_b - I_\tau}\eqno(4.1b)
$$
$$
\lambda_\tau = \lambda_{\tau_G} \mathop{\Pi}_{a}
\zeta^e_a \: e^{-3I_b - 4I_\tau}\eqno(4.1c)
$$
where the subscript $G$ refers to the GUT scale.
The integrals $I_i$ are given by
$$
I_i = \int^{\ln M_G}_{\ln M_S} \left( {\lambda_i (t) \over 4\pi}
\right)^2 dt\eqno(4.2)
$$
and are shown in Figure 2 for $i=t,b,\tau$.
The gauge coupling renormalizations are given by the $\zeta^i_a$ factors, where
$i$ refers to $u,d,e$ flavor and $a=1,2,3$ to the gauge group U(1), SU(2)
or SU(3)
$$
\zeta^i_a = \left({\alpha_G\over \alpha_a}\right)^{c^i_a/2b_a}\eqno(4.3)
$$
where $c^u_a = (13/15, 3, 16/3), \ c^d_a = (7/15, 3, 16/3)$ and $c^e_a =
(27/15, 3, 0)$.
The one loop gauge beta function coefficients are $b_a = (33/5, 1, -3)$.
We remind the reader that $M_S$ is the effective scale of supersymmetry
breaking, and above this scale our theory has the particle content of the
minimal supersymmetric standard model up to mass scale $M_G$.
In this section
we ignore GUT and supersymmetric threshold correction effects, which we expect
to affect our predictions at the level of several percent.

In analytic results given later in the paper the running of the gauge
couplings is treated as follows.
At $\mu = M_Z$ we input $\alpha^{-1}_1 = 58.8$, $\alpha^{-1}_2 = 29.8$ but keep
$\alpha_3(M_Z)$ a free parameter. Except for a special value of $\alpha_3(M_Z)$
(0.108) the three gauge coupling do
not exactly meet at the GUT scale.
This we assume to be due to GUT threshold corrections.
This allows us to show how our predictions vary with $\alpha_3(M_Z)$.
 We assume that the GUT mass relations are valid at a scale $M_G = 2
\times 10^{16} $
 GeV.
This could also be affected by GUT threshold corrections, but is unlikely to
change our predictions by more than a few percent.

The RG scaling of masses of lighter generations are best shown as mass ratios,
as this removes the $\zeta^i_a$ factors
$$
{\lambda_{u,c}\over \lambda_t}  =
   \left( {\lambda_{u,c}\over\lambda_t}\right)_G
 \: e^{3I_t + I_b} \eqno(4.4a)
$$
$$
{\lambda_{d,s}\over \lambda_b}  =
\left( {\lambda_{d,s}\over\lambda_b}\right)_G
 \: e^{I_t + 3I_b} \eqno(4.4b)
$$
$$
{\lambda_{e, \mu}\over \lambda_\tau}  =  \left( {\lambda_{e, \mu}
\over\lambda_\tau}
\right)_G
 \: e^{3I_\tau}. \eqno(4.4c)
$$
We remind the reader that in these formulas the Yukawa couplings
$\lambda_i$
are the eigenvalues of the Yukawa matrices at scale $\mu = M_S$.

The scaling of the KM matrix elements is extremely simple:
$$
V_{cb} = V_{cb_G} \: e^{I_t + I_b}\eqno(4.5)
$$
and identical behavior for $V_{ub}, V_{ts}$, and $V_{td}$.
The CP violating quantity $J$ scales as $V^2_{cb}$, ie
$$
J = J_G \: e^{2I_t + 2I_b}\eqno(4.6)
$$
To the level which we work, the following quantities are RG invariant:\\
 $V_{us},
V_{cd}, V_{ud}, V_{cs}, V_{tb}, \lambda_u/\lambda_c,
\lambda_d/\lambda_s$ and $\lambda_e/\lambda_\mu$.

Finally we must compare our predictions at $M_S$ with parameters
extracted from experiment.
Since $M_S$ is close to the weak scale, we compare the elements of
the KM matrix
at $M_S$ directly with those determined experimentally.
Similarly we take the running top mass to be
$$
m_t \simeq \lambda_t (M_S) {v\over \sqrt{2}} \sin \beta\eqno(4.7)
$$
and the ``physical'' top mass as

$$
M_t \simeq m_t \left(1+ {4\over 3} {\alpha_s(m_t)\over \pi}\right).\eqno(4.8)
$$
Hence we are choosing $M_S$ to be in the neighborhood of the top mass, for
definiteness we take 180 GeV.

The other fermion masses require that we
RG scale the Yukawa parameters below $M_S$.
We define RG scaling parameters $\eta_i$ by
$$
m_i (m_i) \equiv \eta_i m_i (M_S) \eqno(4.9a)
$$
for all quarks and leptons except $u, d$ and $s$ quarks which
have $\eta_i$ defined by
$$
m_i (1 GeV) \equiv \eta_i m_i (M_S) \ i=u,d,s.\eqno(4.9b)
$$
Plots of $\eta_i$ are shown in Figure 3 for $i=u,d,s,c,b$ where the QCD
contribution has been calculated to three loops.
We also include one-loop QED contributions to the $\eta_i$.

We now turn to a discussion of the degree to which each of the four $SO(10)$
operators which contribute to fermion  masses have a unique $SO(10)$ structure.
We first discuss the 33 operator. There are two possibilities for a
renormalizable 33 operator in
$SO(10)$: $16_3\ 10\ 16_3$ and $16_3\ \overline{126}\ 16_3$.
The latter case implies $\lambda_{\tau_G}=3\lambda_{b_G}$, which can only be
turned into a successful $m_b/m_\tau$ relation if $\alpha_s$ is large and if
$m_b$ is less than about 4.0 GeV \footnote{This could, perhaps, be overcome
by large weak-scale radiative corrections to the b quark mass.} \cite{HRS}.
While this case is not excluded, we have chosen to use the operator $16_3\ 10\
16_3$ in this paper, since this is known to work very well \cite{ALS}.
The grand unified boundary condition on the three Yukawa couplings of the
heaviest generation
\footnote{ There are perturbative corrections to this formula.}
, $\lambda_{t_G} = \lambda_{b_G} = \lambda_{\tau_G}$
is reminiscent of that for the three gauge couplings
$g_{1_G} = g_{2_G}=g_{3_G}$. The two free parameters, $A$ and $\tan\beta$, are
determined by $m_b$ and $m_\tau$ via
$$
{m_b\over m_\tau} = {\eta_b\over\eta_\tau} \mathop{\Pi}_{a}
\left({\zeta^d_a\over
\zeta^e_a}\right) \: e^{- I_t -3I_b+3I_\tau}\eqno(4.10)
$$
for $A$ and
$$
m_\tau = {v\over \sqrt{2}} \cos\beta A\ \eta_\tau\mathop{\Pi}_{a} \zeta_a^e \:
e^{-3I_b-4I_\tau}\eqno(4.11)
$$
for $\tan\beta$.
The two predictions for $M_t$ and $\tan\beta$, cannot be given in simple
analytic equations, and are shown in Figures 1a and 1b as a function of the
input $m_b$ for various $\alpha_s$.
In our ``analytic'' analysis for the lighter generation we will read $I_i,
m_t$, and $\tan\beta$ from Figures 1, 2, and 3  and will never
actually use (4.10) or (4.11).
Note that the values of $M_t$ which we predict are larger than those given
by Ananthanarayan, Lazarides and Shafi, partly because they used values of
$\alpha_s$ which would be considered low today.

In appendix 3 we prove that, once the 33 operator has been chosen to be
$16_3\ 10\ 16_3$,
the choice for the 12 entry is unique.
The experimental inputs needed to obtain this result are quite mild:
$$
60 MeV < m_s < 360 MeV
$$
$$
0.2 < m_u/m_d < 1.5
$$
and the experimental value for the Cabibbo angle.
The resulting operator is
$$
O_{12} = 16_1
\left( {45_1\over M}\right)^3
10 \left( {45_1\over M}\right)^3
16_2\eqno(4.12)
$$
which yields symmetric entries, i.e. the 12 and 21 entries are equal in
each mass matrix.

In our scheme the large value of $m_t/m_b$ results from a hierarchy in the
doublet VEVs, tan$\beta \gg 1$.
One might then expect that this hierarchy of VEVs would lead to large values
for both $m_c/m_s$ and $m_u/m_d$. However $m_u/m_d$ is of order unity
and is an order of magnitude or more smaller than
$m_c/m_s$ and $m_t/m_b$.
How does our theory solve this mystery of why $m_u < m_d$ while $m_c \gg m_s$
and $m_t \gg m_b$?
The answer is yet again a Clebsch factor of 3.
Every 45, appearing in $O_{12}$ of equation (4.12)
contributes a factor of 3 to $z_d z'_d/z_u z'_u$.
Hence from the determinant of the up and down Yukawa matrices
we derive
$$
{m_u\over m_d} = \left( {m_t\over m_b}\right)^2
\left( {m_s\over m_c}\right)
\left| {z_uz_u'\over z_dz_d'}\right|
{\eta^2_b \eta_u\eta_c\over \eta_d\eta_s}
\: e^{4(I_t-I_b)}\eqno(4.13)
$$
Thus the naive expectation of $m_u/m_d \simeq (m_t/m_b)^2(m_s/m_c)\simeq 200$
is actually enhanced by $RG$ effects by a factor of about 2-3 to become
of order 400-600.
It is the Clebsch of $|z_uz_u'/z_dz_d'| = 1/3^6 \simeq 10^{-3}$
which reduces $m_u/m_d$ to around 0.6-0.8.
Because the $RG$ enhancement is smallest for small values
of $\alpha_s(M_Z)$ the prediction for
${m_u \over m_d}$ favors smaller values of the strong coupling
constant.

The operator (4.12) involves six suppression factors of
 $\langle 45_1\rangle/M$.
This implies that the scale at which $SO(10)$ is broken to $SU(5)\times U(1)
$ by $\langle 45_1\rangle$ is not much less than the fundamental mass scale
of the theory, $M$.

We have shown that $O_{33}$ and $O_{12}$ are unique, and thus identical in
``22'' models and ``23'' models. To go further we must treat the ``22'' and
``23'' textures separately.

\vskip 9pt
\noindent {\bf (5) THE ``22'' TEXTURE}
\vskip 9pt

In this section we present an analytic treatment of models based on the ``22
texture'' shown in equation $(3.1)$, and so-called because there is an operator
which contributes to the 22 entry of the mass matrices.
We first give the general form of Yukawa matrices of these theories, show how
they may be approximately diagonalized at the GUT scale.
In the last section we showed that $O_{33}$ and $O_{12}$ are unique.
In appendix 4, a numerical search is described which proves that the Clebsch
structure of the 22 entries is also unique.
This allows us to prove that there are just nine possible operators for the
23 entry, and hence nine possible models.
We show how $m_e, m_\mu, m_\tau, m_c, m_b$ and the Cabibbo angle can be used
to accurately determine  the six free parameters which describe the flavor
sector of these theories, and give the resulting eight predictions of the
flavor sector. We find that $m_s/m_d$ disfavors some of the models.
In addition we give the predictions of these models for the kaon CP
impurity parameter $\epsilon$, for $B^0\overline{B^0}$ mixing and for the
CP violating asymmetries in neutral $B$ meson decay.

The GUT scale Yukawa matrices which follow from the theories defined by the
four operators of equation $(3.1) $ can be written in the form:
$$
\bf{\lambda_i} =\pmatrix{ 0&z'_iC&0\cr
z_iC&y_iE e^{i \tilde{\phi}}&x'_iB\cr
0&x_iB&A}\eqno(5.1)
$$
where $i = u,d,e$. In this paper we define the coupling matrices with the
doublet fields to the right of the matrices, so $x_i$ are relevant for
$V_{cb}$.

Notice that $A,B,E,C$ are all dimensionless: they are the original operator
coefficients multiplied by the relevant factor of \ul{45} vacuum expectation
values.
Thus $A \gg B, E \gg C$.
Phase redefinitions on the matter multiplets have been arranged so that the
single physical phase $\tilde{\phi}$ appears in the 22 entry.
The Clebsch factors from the \ul{45} VEVs are parameterized as $x_i, x'_i,
y_i, z_i$ and $z'_i$ and for a given model  can be obtained from Table 1.

Approximate diagonal forms for these matrices at the GUT scale can be
obtained by the following sequence of transformations

i) In the heavy $2\times 2$ sector rotate the left-handed fermions by
angles $\alpha_i = x_i B/A$, and the right-handed ones by
$\alpha'_i = x'_i B/A$.

ii) Write the resulting 22 entry as $y_iE_ie^{i\phi_i}$ and rotate the left
and right-handed lightest generation fermions by a phase factor
$e^{i\phi_i}$.

(iii) Diagonalize the light $2\times 2$ sector by rotations
$\beta_i = z_iC/y_iE_i$ and $\beta'_i = z'_iC/y_iE_i$ on the left and right-
handed fermions respectively.

Let $\overline{\theta}_3 = - \alpha_u,\ \overline{\theta}_4 = -\alpha_d,\
\theta_2 = -\beta_u$ and $\theta_1 = \beta_d$, then the transformations in the
left-handed up and down sectors are:
$$
V_u = \pmatrix{ c_2&s_2&0\cr
-s_2&c_2&0\cr
0&0&1}
\pmatrix{ e^{i\phi_u}&0&0\cr
0&1&0\cr
0&0&1}
\pmatrix{ 1&0&0\cr
0&\overline{c_3}&\overline{s_3}\cr
0&-\overline{s_3}&\overline{c_3}}\eqno(5.2a)
$$
and
$$
V_d = \pmatrix{ c_1&-s_1&0\cr
s_1&c_1&0\cr
0&0&1}
\pmatrix{ e^{i\phi_d}&0&0\cr
0&1&0\cr
0&0&1}
\pmatrix{ 1&0&0\cr
0&\overline{c_4}&\overline{s_4}\cr
0&-\overline{s_4}&\overline{c_4}}\eqno(5.2b)
$$
Setting $\theta_3 = \overline{\theta_3} -\overline{\theta_4}$
and $\phi = \phi_u - \phi_d$, the KM matrix
takes the form:
$$
V = V_u V^{\dagger}_d =
\pmatrix{ c_1c_2-s_1s_2e^{-i\phi}&s_1+c_1s_2e^{-i\phi}&s_2s_3\cr
-c_1s_2-s_1e^{-i\phi}& c_1c_2c_2 e^{-i\phi}-s_1s_2& c_2s_3\cr
s_1s_3&-c_1s_3& c_3e^{i\phi} }
\eqno(5.3)
$$
which is identical in form to the KM matrix obtained in a previous framework
\cite{DHR}.  The angles $\theta_1$ and $\theta_2$ are related to quark
Yukawa couplings and the Clebsch factors $z_i$ and $z'_i$ by

$$
   s_1 = \sqrt{ |{z_d \over z_d'}| } \sqrt{ {\lambda_d \over \lambda_s }}
$$
$$
   s_2 = \sqrt{|{z_u \over z_u'}|} \sqrt{ {\lambda_u \over \lambda_c }}
$$
Using results from the last section it is straightforward to see that
$\theta_1, \theta_2$ and $\phi$ are RG invariants, while $s_3$ scales as
$V_{cb}$.

We now turn to a discussion of the 22 and 23
operators, which generate $V_{cb_G}, \lambda_{c_G}, \lambda_{s_G}$ and
$\lambda_{\mu_G}$.
The low energy experimental values of $V_{cb} = .044 \pm .006, \ m_c = 1.22
\pm .05$ GeV, $m_s = 175 \pm 55$ MeV and $m_\mu = 106$ MeV can be run up to
the GUT scale with the RG equations yielding
$$
\eqalignno{
V_{cb_G} &= 0.040 \ {\hbox{to}} \ 0.024 &(5.4a)\cr
\left( {\lambda_c\over \lambda_t}\right)_G &= 0.0030\ {\hbox{to}}\
0.0012&(5.4b)\cr
\left( {\lambda_s\over \lambda_b}\right)_G&= 0.025 \ {\hbox{to}}
\ 0.009&(5.4c)
\cr
\left({\lambda_\mu\over\lambda_\tau}\right)_G &= 0.048\  {\hbox{to}}\
0.035&(5.4d)
\cr}
$$
where the quoted ranges correspond to $\alpha_s(M_Z) = 0.11$ to 0.13
respectively.
In the cases of (5.4 b,c,d) we have also used our prediction for $m_t$
and inputs $m_b =
4.15 \pm 0.1$ GeV and $m_\tau = 1.78$ GeV.
A crucial feature concerning the magnitude of these four quantities is:
$$
V^2_{cb_G} \simeq \left| {\lambda_c\over \lambda_t}\right|_G  \ll
\left| {\lambda_\mu\over \lambda_\tau}\right|_G\simeq 3
\left| {\lambda_s\over\lambda_b}\right|_G.\eqno(5.4e)
$$
The theory can naturally account for this division into two small parameters
and two large ones if the 23 operator generates the small parameters and the
22 operator the large ones.
This requires a 22 operator which gives $| y_u | : | y_d | : | y_e | =\:
< 1/3: 1 : 3$.
A study of all operators at dimension 5 and 6 shows there to be a unique
$y_i$ which satisfies this
$$
y_u : y_d : y_e = 0 : 1 : 3.\eqno(5.5)
$$
This is the form familiar from the Georgi-Jarlskog texture.
There are six operators which give such Clebsch ratios
$$
\eqalignno{
16_2 \ 45_1  &10 \ {45_{B-L} \over 45_1} \ 16_2&(5.6a)\cr
16_2 {1 \over \ 45_1}  &10 \ 45_{B-L} \ 16_2&(5.6b)\cr
16_2 \ 45_1  &10 \ 45_{B-L} \ 16_2&(5.6c)\cr
16_2 &10 \ {45_{B-L}\over 45_1}\  16_2&(5.6d)\cr
16_2 &10 \ 45_1 \ 45_{B-L} \ 16_2 &(5.6e)\cr
16_2 &10 \ {45_{B-L} \over 45_1^2}\ 16_2 & (5.6f)\cr}
$$
Since they lead to the same $y_u : y_d: y_e$, the operators lead to
identical predictions.
All six operators require a 45$_{B-L}$ in addition to the 45$_1$, which is
needed both here and for $O_{12}$.

The 23 operator generates not only
$$
V_{cb_G} = (x_d - x_u) {B\over A} \eqno(5.7)
$$
but is also  entirely responsible for generating $\lambda_{c_G}$
$$
\left({\lambda_c\over \lambda_t}\right)_G = | x_ux_u' | \left({B\over A}\right)
^2.\eqno(5.8)
$$
Since $B/A$ is determined by the charm quark mass, the Clebsch combinations
$x_ux_u'$ and $x_u - x_d$ cannot both be independently probed.
Eliminating $B/A$ gives
$$
V_{cb_G} = \chi \sqrt{ \left| {\lambda_c\over \lambda_t}\right|_G}\eqno(5.9)
$$
where the Clebsch combination which can be measured is given by
$$
\chi \equiv {|x_u-x_d|\over \sqrt{|x_ux_u'|}} \eqno(5.10)
$$
and the sign of B/A was chosen to make $V_{cb}$ positive.
Experimental values for $V_{cb}$ and $m_c$, together with the predicted
value of $m_t (\alpha_s)$ imply that allowed values of $\chi$ must
fall in the range
$$
0.55 < \chi < 0.92\eqno(5.11)
$$
where the larger values of $\chi$ tend to result from large $V_{cb}$ and large
$\alpha_s$.

A search of all operators of dimension 5 and 6 shows that only three values
of $\chi$ occur in this range: $\chi = 2/3, 5/6, 8/9$,
resulting from the operators:
$$
\eqalignno{
\chi = 2/3 &\cr
(1)&16_2\ 45_{24} \ 10 {1\over 45_1} \ 16_3\cr
(2)&16_2 \ 45_{24} \ 10\ {45_{B-L}\over 45_1} \ 16_3\cr
(3)&16_2 \ {45_{24}\over 45_1} \ 10 \ {1\over 45_1} \ 16_3\cr
(4)&16_2 \ {45_{24}\over 45_1} \ 10 \ {45_{B-L}\over45_1} 16_3&(5.12a)\cr}
$$
$$
\eqalignno{
\chi = 5/6 &\cr
(5)&16_2 \ 45_{24} \ 10 \ {45_{24}\over 45_1} \ 16_3\cr
(6)&16_2 \ {45_{24}\over 45_1} \ 10 \ {45_{24}\over 45_1} \ 16_3&(5.12b)\cr}
$$
$$
\eqalignno{
\chi= 8/9&\cr
(7)&16_2 \ 10 \ {1\over 45_1^2} \ 16_3\cr
(8)&16_2 \ 10 \ {45_{B-L}\over 45^2_1} \ 16_3\cr
(9)&16_2 \ 10 \ {45_{B-L}^2\over 45^2_1} \ 16_3&(5.12c)\cr}
$$
In equations (5.12) we label the operators (1) - (9), and will use these
numbers also to denote the corresponding models.
If slightly larger values of $V_{cb}$ and $\alpha_s$ are accepted, the case
$\chi=1$ is also allowed\footnote{We have specifically allowed for the
possibility $v_{10} > v_5$ in order to obtain models with $\chi < 1$.}. This
case has been studied previously \cite{ADHR} and arises when $\lambda_{tG} =
\lambda_{bG}$ is imposed on the Georgi-Jarlskog texture, which has $x_d =0$
and $x'_u = x_u$ and therefore $\chi=1$. It is interesting to note that this
can be quite simply obtained by the operator
$$
   16_2 \; 10 \;  45_1 45_Y \; 16_3     \eqno(5.13)
$$
At first sight the Georgi-Jarlskog texture seems to be in conflict with the
idea of up-down $SO(10)$ mass relations: $U_{22} =0$ but $D_{22} \neq 0$,
and $U_{23} \neq 0$ but $D_{23} =0$.
We find that this is not the case.
The Georgi-Jarlskog predictions
result with the 23 operator of equation (5.13) and any of the 22 operators
of (5.6).
The condition $\lambda_{t_G} = \lambda_{b_G}$, which was not part of the
original Georgi-Jarlskog scheme, implies that $\chi=1$ results only if
$\alpha_s$ and $V_{cb}$ are uncomfortably large, so we do not consider it
further.

We have written down the most general set of 22 and 23 operators which
follow from the assumption that $\lambda_{\mu_G}$ and $\lambda_{s_G}$ are
dominated by the 22 operator, while $\lambda_{c_G}$ and $V_{cb_G}$ have
contributions only from the 23 operator.
While we have argued that this is a natural division, it is not obvious that
it is a necessary one.
In section 6 we mention a numerical search for all possible
models with the texture of (5.1) with the $z_i, z_i'$ Clebsches coming
from operator (4.12), which we have already proved is unique.
The result of this numerical search is just the models described by the 22
operators of equation (5.6) and the 23 operators of equation (5.12).
These well-motivated models are the unique ones of this texture.

Thus, without any loss of generality, we are now able to write the GUT-scale
Yukawa matrices which result from this texture as:
$$
{\bf U} = \pmatrix{ 0&-{1\over 27}C&0\cr
-{1\over 27}C&0&x'_uB\cr
0&x_uB&A}
$$
$$
{\bf D} = \pmatrix{0&C&0\cr
C&Ee^{i \tilde{\phi}}&x'_dB\cr
0&x_dB&A}
$$
$$
{\bf E} =
\pmatrix{ 0&C&0\cr
C&3Ee^{i \tilde{\phi}}&x'_eB\cr
0&x_eB&A }
\eqno(5.14)
 $$
The unique $z_i, z_i', y_i, y'_i$ Clebsches have been shown explicitly,
while the $x_i, x_i'$ parameters allow for the possible
Clebsches which follow from the 23 operators of equation (5.12).

The fermion masses and mixing angles depend on six free parameters:
$A,B,C,E, \tilde{\phi}$ and $\tan\beta$.
We will now give the six equations that fix these parameters in terms of $m_e,
m_\mu, m_\tau, m_c, m_b$ and $\sin\theta_c$.
$A$ and $\tan\beta$ are given in terms of $m_b$ and $m_\tau$ in equations
(4.10) and (4.11).
$B/A$ is determined from $m_c$ via the equation
$$
{m_c\over m_t} = \eta_c |x_ux'_u| \left({B\over A}\right)^2 \: e^{3I_t +I_b}
\eqno(5.15)
$$
with $m_t$ determined from Figure 1a and equation (4.8).
Thus $B/A$ actually depends on two inputs $m_c$ and $m_b$, and on
$\alpha_s$.
The ratio $C/A$ is obtained by taking the determinant of the lepton Yukawa
matrix of equation 5.14, and dividing by $\lambda^3_{t_G}$:
$$
{m_e m_\mu\over m_\tau^2} = {\eta_e\eta_\mu\over \eta_\tau^2}
 \left({C\over
A}\right)^2 \: e^{6I_\tau} \eqno(5.16)
$$

We are left with the determination of $E$ and $\phi$ from $m_\mu$ and
$\sin\theta_c$.
Diagonalization of the heavy $2\times 2$ sectors of the {\bf D} and {\bf E}
matrices gives
$$
\left( {\lambda_\mu\over \lambda_\tau}\right)_G = 3 {E\over A} \
\sqrt{ 1 -2 \delta_e \cos \tilde{\phi} + \delta^2_e}\eqno(5.17a)
$$
$$
\eqalignno{
\left( {\lambda_s\over\lambda_b}\right)_G &= {E\over A} \
\sqrt{1 -2 \delta_d \cos \tilde{\phi} + \delta^2_d}\cr
&\simeq {E\over A} \ ( 1-\delta_d \cos \tilde{\phi})&(5.17b)\cr}
$$
where the last step involves the approximation $\abs{\delta_d} \ll 1$,
valid in all 9 of our models.
Here
$$
\delta_e = {x_ex'_e\over 3} {B^2\over AE} \eqno(5.18a)
$$
$$
\delta_d = x_dx_d' {B^2\over AE}. \eqno(5.18b)
$$
Scaling (5.17a) to low energies gives
$$
{m_\mu\over m_\tau} = 3\abs{{E\over A}} \sqrt{ 1 -2 \delta_e\cos
\tilde{\phi} +
\delta_e^2}\ {\eta_\mu\over \eta_\tau} \ e^{3I_\tau}\eqno(5.19)
$$
which provides one equation for $E$ and $\tilde{\phi}$.

The phase $\tilde{\phi}$ of the Yukawa matrices differs in general from the
phase $\phi$ of the CKM matrix. In the ``22" texture $\phi = \phi_u - \phi_d =
- \phi_d$ and the difference between $\tilde{\phi}$ and $\phi_d$ is small:
$\tilde{\phi} - \phi_d \approx \delta_d$. In all nine models of interest
$\delta_d$ is less than 0.1, and in most less than 0.01, hence in the rest of
this section we take $\tilde{\phi} = - \phi$. A relation to determine $\phi$
follows from:
$$
\sin \theta_c = |V_{us}| = | s_1+ c_1 s_2 e^{-i\phi}|\eqno(5.20)
$$
 so that $\phi$ is determined from
$$
\cos \phi = {\sin^2\theta_c-s^2_1 - s_2^2 \over 2s_1s_2}\eqno(5.21)
$$
where the angles $\theta_1$ and $\theta_2$ are given by
$$
\tan \theta_1= {C\over \lambda_{s_G}} \simeq {C\over E} (1 + \delta_d
\cos\phi).\eqno(5.22)
$$
$$
s_2 = {C\over 27 \lambda_{c_G}} \simeq - { 1 \over 27 x_ux_u'} \ {AC\over B^2}
\eqno(5.23)
$$
The ratios $B/A, C/A$ and $E/A$ necessary to evaluate $s_1$ and $s_2$ are
obtained from (5.15), (5.16) and (5.19). The sign of $B/A$ has been chosen to
make $s_3 > 0$, and we now choose the sign of $C/A$ and $E/A$ to make $s_1$ and
$s_2$ positive.
Thus, without loss of generality, the angles $\theta_1$, $\theta_2$ and
$\theta_3$ are all taken to lie in the first quadrant.
Hence we have
$$
\tan \theta_1\simeq 3 \sqrt{{m_e\over m_\mu}} \sqrt{{\eta_\mu\over \eta_e}}
(1-\delta\cos \phi)\eqno(5.24)
$$
where we have defined $\delta$ by
$$
1- \delta \cos \phi \equiv \sqrt{ 1-2\delta_e \cos \phi + \delta_e^2} \
(1+ \delta_d \cos\phi).\eqno(5.25)
$$
In most of our models $|\delta_e| \ll 1$, in which case $\delta \simeq
\delta_e - \delta_d$.
Equation (5.24) is alternatively written as
$$
s_1\simeq .196 (1- \delta \cos \phi)\eqno(5.26)
$$
Inserting $B/A$ and $E/A$ into (5.18a) gives
$$
\eqalignno{
\delta_e = &- {x_e x_e'\over x_u x_u'}
\sqrt{1-2\delta_e\cos\phi
 + \delta_e^2}\
{m_\tau\over m_\mu}\ {m_c\over m_t} \ {\eta_\mu\over \eta_\tau\eta_c}
e^{-3I_t - I_b + 3I_\tau}&(5.27a)\cr}
$$
and
$$
\delta_d = {3x_dx_d'\over x_ex_e'} \ \delta_e. \eqno(5.27b)
$$
The signs of $\delta_e$ and $\delta_d$ will be crucial: in (5.27) it is
understood that $\theta_1$ and $\theta_2$ have been chosen in the first
quadrant.
Finally, using the determined values for $C/A$ and $B/A$ in (5.23) gives
$$
s_2 = {1\over 27} \ {m^{1/2}_e m^{1/2}_\mu m_t\over m_\tau m_c} \
{\eta_c\eta_\tau\over \eta^{1/2}_e \eta^{1/2}_\mu} \ e^{3I_t + I_b - 3I_\tau}
\eqno(5.28)
$$
The analytic determination of $\cos \phi$ in a particular model is not
completely straightforward.
This is because $\cos \phi$ occurs in both (5.26) and (5.27a) so the equation
(5.21) becomes a polynomial of high order in $\cos \phi$.
The analysis simplifies considerably for the
models with $\abs{\delta_e} \ll 1$,
since then the square root factor in (5.27a) may be neglected in estimating
$\delta_e$.

This concludes the parameter determination:

$\bullet$ $A$ from $m_b/m_\tau$ via (4.10)

$\bullet$ tan $\beta$ from $m_\tau$ via (4.11)

$\bullet$ $B/A$ from $m_c/m_t$ via (5.15)

$\bullet$ $C/A$ from $m_em_\mu/m^2_\tau$ via (5.16)

$\bullet$ $E/A$ from $m_\mu/m_\tau$ via (5.19)

$\bullet$ $\phi$ from $\sin \theta_c$ via (5.21).
\vskip 9pt

Before proceeding to the eight predictions, we show why
the prediction for $m_s/m_d$ disfavors some of the nine models.
{}From the determinant of {\bf D} and {\bf E} we find
$$
{m_s\over m_d} \left( 1- {m_d\over m_s}\right)^2 = {1\over 9} {m_\mu\over m_e}
\left( 1 - {m_e\over m_\mu}\right)^2 {\eta_\mu\over \eta_e} {1\over (1-\delta
\cos\phi)^2}\eqno(5.29)
$$
where $1-\delta \cos \phi$ is given in (5.25).
This reproduces one of the predictions of Georgi and Jarlskog
\cite{GJ} in the limit
that $\delta_{e, d} \to 0$: $m_s/m_d = 25.15$.
Since this is a high value for $m_s/m_d$, a very interesting feature of the
present models is that this number is modified by $\delta_{e,d}$.
We consider models with $m_s/m_d$ larger than 25.15 to be disfavored.
In general the deviation of $m_s/m_d$ from 25.15 will depend on the
numerical  value of the inputs $m_c$, $m_b$, and $\alpha_s$.  This
dependence, and the quantitative value of $m_s/m_d$ will be discussed
in section 6.  However the qualitative behavior of $m_s/m_d$ can be
understood from our analytic formulas.

To understand the qualitative deviation of $m_s/m_d$,
the crucial question is the signs of $\delta$ and $\cos \phi$ for the nine
models.
The signs of $\delta_e$ and $\delta_d$ are determined by the signs of
$x_ex_e'/x_ux'_u $ and $x_dx_d'/x_ux_u'$ which are listed in Table 2.
\vskip 9pt
\centerline {\bf Table 2}

\vskip 9pt

\begin{center}
\begin{tabular}{|c|c|c|c|c|c|c|c|c|c|}
\hline
&1&2&3&4&5&6&7&8&9\\
\hline
${x_ex_e'\over x_ux_u'}$&$-1.5$&$-$13.5&1/2&4.5&$-6.75$&2.25&1/9&1&9\\
\hline
${x_dx_d'\over x_ux_u'}$&1/6&1/6&$-$1/18&$-$1/18
&$-$1/12&1/36&1/9&1/9&1/9\\
\hline
$c={3x_dx_d'-x_ex_e' \over x_ux_u'}$&2&14&$-$2/3
&$-$4.7&6.5&$-$2.2&2/9&$-$2/3&$-$8.7\\
\hline
$m_s\over m_d$&26.7&$D$&24.8&23.6&$D$&24.2&25.3&24.8&\\
\hline
$\chi$&2/3&2/3&2/3&2/3&5/6&5/6&8/9&8/9&8/9\\
\hline
$\cos \phi$&0.35& &0.24&0.16& &0.20&0.27&0.24&\\
\hline
$\sin \phi$&0.94& &0.97&0.99& &0.98&0.96&0.97&\\
\hline
\end{tabular}
\end{center}
Using (5.29), one can rewrite (5.27a) approximately as:
$$
\delta_e \simeq - 0.043 {x_ex_e'\over x_ux_u'} {m_c\over 1.27 GeV}
\ {180 GeV\over
m_t} \ \sqrt{ {23 \over m_s/m_d}} \ (1 \mp 0.05)\eqno(5.30)
$$
for $\alpha_s(M_Z) = 0.115 \pm 0.005$.
Thus for many models $\abs{\delta_e}$ and $\abs{\delta_d}$ are much less than
unity, in which case $\delta $ appearing in (5.29) is given by $\delta \simeq
\delta_e - \delta_d$.
The sign of $\delta$ is then given by the sign of $ c = {3x_dx_d'
- x_ex_e'\over x_ux_u'}$,
which is also listed in Table 2.
Consider the models with $\delta$ positive: if $\cos \phi$ is also positive
$m_s/m_d$ will be increased above 25.15 and the models disfavored.
For models 1 and 7, $\delta_{e, d} \ll 1$ and
$\delta \simeq \delta_e - \delta_d$ is positive.
Because $\delta \ll 1, $ (5.21) determines $\cos \phi$ to be positive: these
two models do indeed give values of $m_s/m_d$ larger than 25.15.
The values of $m_s/m_d$, as well as $\cos \phi$ and $\sin \phi$
are shown in the table for $\alpha_s (M_Z) = 0.115$. Note that for the models
where c is not large, eq. (5.21) can be approximated as
$ \cos \phi \simeq 0.26/ (1 - 0.13 c)$; where we have used
$\alpha_s(M_Z)=0.115$
and $m_t=180$ GeV.

The other two models with $c={3x_dx_d- x_ex_e'\over x_ux_u'} >0$
are models 2 and 5.
In these cases $|\delta_e|$ is not much less than unity, so the analysis is
more
complicated.
However, it can be shown that $s_1 > .196$  (necessary for $m_s/m_d < 25.15$)
can only occur for a very narrow range of $\delta_e$ near 0.40.
Using equation (5.30) the central values of $\delta_e$ for models 2 and 5 are
0.59 and 0.30.
To obtain an acceptable solution with $\delta_e$ near 0.40, parameters must be
chosen far from their central values.
While not excluded, we find these two models to be highly disfavored:
in table 2 we write ``$D$'' for disfavored.
However, if extreme values of $\alpha_s$ are used in these theories, $m_s/m_d$
can vary over the entire acceptable range.

Models 3, 6, and 8 all have $|\delta_e| \ll 1$ so that
$\delta \simeq \delta_e - \delta_d$ and is negative.
Furthermore, since $\cos \phi$ is positive for these three models, $m_s/m_d$
is reduced slightly below 25.15.
For models 4 and 9 the small $|\delta_e|$ approximation is no longer valid,
nevertheless one finds that in these models $m_s/m_d$ is decreased even
further,
so that, from the viewpoint of $m_s/m_d$, these are the preferred models.
However, in comparing with data it is important to study predictions for
$m_u/m_d$ simultaneously with $m_s/ m_d$, and this will be done in the next
section.
{}From the numerical results of Figure 5 one can see that $m_s/m_d$ decreases
in the sequence of models : 3 and 8, 6, 4, 9. This can be understood as
successively larger deviations from the Georgi-Jarlskog result due to the
Clebsch factor $c = {3x_dx_d' - x_ex_e'\over x_ux_u'}$ shown in table 2.

We now give analytic formulas for the eight predictions.
The top mass $m_t$ and $\beta$ can be predicted from the analysis of the
heaviest generation.
With $A$ and therefore $I_i$ determined from (4.10):
$$
\cos \beta = { \sqrt{2}m_\tau\over Av\Pi_a \zeta^e_a \eta_\tau} \
e^{3I_b+4I_\tau}\eqno(5.31)
$$
$$
m_t = \mathop{\Pi}_{a} \left( {\zeta^u_a\over \zeta^e_a}\right)
\tan\beta \ m_\tau\:  e^{-6I_t + 2I_b + 4I_\tau}.\eqno(5.32)
$$
$RG$ scaling (5.9) to low energies gives
$$
V_{cb} = \chi \sqrt{{m_c \over m_t}} {1 \over \sqrt{\eta_c}}
e^{{1 \over 2}(I_b - I_t)}
\eqno(5.33)
$$
where $\chi$ is given in (5.10).
This result illustrates how our predictions have the form:
$$
\pmatrix{\hbox{Predicted}\cr\hbox{Quantity}} =
\pmatrix{\hbox{GUT}\cr \hbox{Clebsch}}
\pmatrix{\hbox{Input}\cr\hbox{Quantity}}
\pmatrix{\hbox{RG}\cr \hbox{factor}}
$$
The strange mass is obtained from (5.17b)
$$
m_s \simeq {1\over 3} m_b {m_\mu\over m_\tau}
 { 1 \over 1 - \delta \cos \phi} {\eta_s\over \eta_b}
{ \eta_\tau\over \eta_\mu}\: e^{3I_b + I_t - 3I_\tau}\eqno(5.34)
$$
which can be simplified using (4.11) for $m_b/m_\tau$.
The prediction for $m_s/m_d$ follows
from the determinant of {\bf D} and {\bf E} and is given in (5.29).
The prediction for $m_u/m_d$ follows from the determinantal relation (4.13)
with $m_s$ from (5.34):
$$
{m_u\over m_d} \simeq {1\over 3^7} {m_\mu\over m_\tau} { m^2_t\over m_cm_b}
 { 1 \over 1 - \delta \cos \phi} \eta_b\eta_c {\eta_\tau \over \eta_\mu}
\:e^{5I_t-I_b -3I_\tau}\eqno(5.35)
$$

{}From the form of the KM matrix of equation (4.3) one finds $V_{ub}/V_{cb} =
s_2$,
and using (5.28) this immediately gives
$$
{V_{ub}\over V_{cb}} \simeq {1\over 27} {m_e^{1/2}m_\mu^{1/2}\over m_\tau}
{m_t\over m_c} {\eta_c\eta_\tau\over \eta^{1/2}_e \eta^{1/2}_\mu}\: e^{
3I_t + I_b - 3I_\tau}\eqno(5.36)
$$
The final prediction is for the amount of CP violation in the KM matrix,
which we choose to specify via the rephase invariant quantity $J$\cite{J}
$$
\eqalignno{
J &\equiv Im V_{ud} V_{tb} V^*_{td} V^*_{ub} \simeq s_1s_2s_3^2s_\phi\cr
  &\simeq {\chi^2\over 9} {m_e\over m_\tau} (1-\delta
\cos \phi) s_\phi {\eta_\tau\over
\eta_e}\: e^{2I_t + 2I_b -3I_\tau}&(5.37)\cr}
$$
where $\phi$ is obtained from (5.21).
Since $s_\phi$ is determined to be near unity, the value of $J$ can be
obtained to within a factor of 2 from $J \simeq {\chi^2 \over 9}
{m_e\over m_\tau} \simeq 3 \times 10^{-5} \chi^2$.

In summary our eight predictions for $\beta, m_t, V_{cb}, m_s, m_s/m_d,
m_u/m_d, V_{ub}/V_{cb}$ and $J$ are given in (5.29) and (5.31)-(5.37).
An interesting feature of these predictions, is that of the six Clebsches
which appear in the Yukawa matrices ($x_u, x'_u, x_d, x'_d, x_e$ and $x'_e$)
only three linear combinations occur in physical observables:
$
\chi = |x_u -x_d|/\sqrt{|x_ux'_u|}$ and
$x_ex_e' /x_ux'_u$ and $x_dx_d' /x_ux'_u$ which occur in $\delta_e$ and
$\delta_d$.
The combination $\chi$ occurs in the $V_{cb}$ prediction and we have already
seen that this limits us to consider only theories with $\chi=2/3, 5/6,$ and
$8/9$.
\vskip 9pt

In comparing the nine models, it is useful to recall that
the $m_t$ and $\tan \beta$ predictions are essentially universal.
The distinctions between the models appear in the
other predictions, which we now evaluate for inputs of
$m_b = 4.25$  GeV, $m_c = 1.27$ GeV and $\alpha_s(M_Z) = 0.115 \pm .005$.
In this case we find
$$
V_{cb} \simeq .059 \chi \sqrt{{180 GeV \over m_t}} \ (1 \mp .03)\eqno(5.38)
$$
$$
{V_{ub}\over V_{cb}} \simeq .063\ {m_t \over 180 GeV} \ (1 \pm 0.1)\eqno(5.39)
$$
where the range corresponds to
$\alpha_s (M_Z) = 0.115 \pm 0.005$. Note that the predicted values
for the running mass $m_t = m_t(m_t)$
should be used in these equations and
in the remainder of this section. For the relevant
$m_b$ and $\alpha_s$ the predicted top pole mass,
$M_t$, should be read from Figure 1a,
and converted to the running mass using eq. (4.8).

Predictions for $m_s, m_u/m_d$ and J can be obtained from eqs. (5.34), (5.35)
and (5.37):
$$
m_s \simeq (168  \pm 19) \sqrt{ {m_s/m_d\over 23}} MeV\eqno(5.40)
$$
$$
{m_u\over m_d} \simeq 0.71\ \left( m_t \over 180 GeV \right)^2
\sqrt{ {m_s/m_d \over 23}}(1\pm 0.13)\eqno(5.41)
$$
$$
J \simeq 4.6 \times 10^{-5} \chi^2
\sqrt{ {23\over m_s/m_d}} \sin \phi (1 \pm 0.03) \eqno(5.42)
$$
for $\alpha_s(M_Z) = 0.115\pm 0.005$.
Note that eq. (5.29) has been used to
express $(1-\delta \cos \phi)$ in terms of $m_s/m_d$.
{}From Table 2 it can be seen that $\sin \phi \simeq 1$.

Finally we give approximate formulas for CP violation in $K$ and $B$
meson processes, and for $B^0\overline{B}^0$ mixing.
Using the same central values as above, we find the kaon CP impurity
parameter $\epsilon$ is given by
$$
|\epsilon | \simeq 2.26 \ 10^{-3} \left({\hat{B}_K\over 0.51}\right)
\left({\chi\over 5/6}\right)^4 \
\left( {23\over m_s/m_d}\right)^{3/2} \ \left({m_t \over 180 GeV}\right)^{1/2}
(1 \mp 0.03)\eqno(5.43)
$$
for $\alpha_s (M_Z) = 0.115 \pm 0.005$.
A lattice calculation gives $\hat{B}_K = 0.72 \pm 0.06$ \cite{S}, where the
error includes only the uncertainty due to the continuum extrapolation.
Other calculational approximations, such as the quenched
approximation, lead to additional uncertainties.
For $B^0_d\overline{B}^0_d$
mixing we find
$$
x_d \equiv {\Delta m\over \Gamma} \simeq
0.67 {\tau_B \over 1.28 ps}  \left({\sqrt{B}f_B\over 175 MeV}\right)^2
 \left({\chi\over 5/6}\right)^2 \ {23\over m_s/m_d} \
\left({m_t \over 180 GeV}\right)^{1/2} (1\mp 0.06)\eqno(5.44)
$$
showing that large $\chi$ prefers smaller $f_B$
(which is normalized such that $f_\pi = 135$ MeV).

The CP violating parameters sin 2$\alpha$ and sin 2$\beta$ measured in $B^0 \to
\pi^+\pi^-$ and $B^0\to \psi K_s$ are given in our models by
$$
\sin 2\alpha = - 2 \cos \phi \sin \phi\eqno(5.45a)
$$
$$
\sin 2\beta = {2c_1s_1s_2\over s^2_c} \sin \phi \left( 1 + {c_1s_2\over s_1 }
\cos \phi\right).\eqno(5.45b)
$$
{}From Table 2 it can be seen that the six models
where c is not large have $|\cos \phi| < 0.35$.
In this case the factor ${c_1s_2\over s_1} \cos\phi$ is less than 0.1 and can
be neglected in (5.45b).
Since these models all have sin $\phi \simeq 1$ we find $\sin 2\beta =
2s_1s_2/s_c^2$.
Using $s_1$ from eq. (5.26) and $s_2 \simeq 0.063 \ (1\pm 0.1)$ we
find that these models all lead to essentially the same $\beta$:
$$
\sin 2\beta \simeq 0.54\ \sqrt{ {23 \over m_s/m_d}} (1 \pm 0.15).\eqno(5.46a)
$$
On the other hand these six models do have variation in $\cos \phi$,
as shown in Table 2, so they can be
distinguished by $\sin 2\alpha$:
$$
\sin 2\alpha \simeq - (0.66, 0.47, 0.32, 0.40, 0.52, 0.47) \eqno(5.46b)
$$
for models 1,3,4,6,7 and 8 respectively.

These numbers, and the predictions of table 2, are for $\alpha_s(M_Z)=0.115$
and $m_t = 180$ GeV. The full predictions for these models, and also for models
2,5 and 9 where c is large, will be given in the next section. In Appendix 4 we
show that the predictions of this section survive, with certain modifications,
even if there are large supersymmetric threshold corrections to $\lambda_b$
which give a lighter top quark. In appendix 5 we discuss the
extent to which these predictions are affected by GUT threshold
corrections, including relaxation of the assumption that both Higgs doublets
lie in a single 10.

\vskip 9pt
\noindent{\bf (6) A QUANTITATIVE STUDY OF THE ``22'' TEXTURE}
\vskip 9pt

 In this section we present a quantitative analysis of the eight predictions
in each of the models with ``22" texture.  To obtain the numbers quoted
in this section we numerically integrated the full two-loop renormalization
group equations from the grand unification scale down to the weak scale.
At the grand unified scale, a small threshold correction was applied
to the strong coupling constant in order to obtain a prescribed
value of $\alpha_s(M_Z)$.
At the weak scale, one loop matching corrections were applied to
account for
a degenerate spectrum of superpartners.  For definiteness sake, we
show our predictions for $M_{SUSY} = M_{WEAK} = 180 GeV$, however,
we have varied both these scales separately and our results do not
change by more than a few percent.  At the weak scale, the Yukawa
couplings were diagonalized and the mixing angles were determined.
The fermion masses obtained at the weak scale were subsequently evolved
to three loops in QCD and two loops in QED from the weak scale down to
the larger of either the particles mass or $1$ GeV.

   In addition to the numerical evolution described above we have
performed a numerical search to discover all possible models having the
``22" texture.  Using the uniquely viable operators
$O_{12}$ and $O_{33}$
discussed in section 4, we searched through all possible pairings
of operators $O_{22}$ and $O_{23}$ up to dimension 6.
In this search, we demanded $m_s = 199 \pm 66 MeV$, $0.5 <\chi < 1$,
where $V_{cb} = \chi \sqrt{m_c/m_t}$, $ 0.03 < V_{ub}/V_{cb} < .1$,
$ 17 <m_s/m_d < 26$, and $.2 <m_u/m_d < .7$.   All of the models
which passed these constraints simultaneously had a ``22" operator
with a 0:1:3 Clebsch, thus establishing the uniqueness of
the $O_{22}$ operators in equation (5.6). Furthermore, the search resulted in
nine acceptable ``23" operators, which are precisely those given in equation
(5.12), and we now turn to the predictions of these
nine different ``22" models.

Because the large hierarchy within the third generation is explained by the
dynamical factor $\tan\beta$, and because an acceptable $m_b/m_\tau$
requires an ${\cal O} (1)$ top quark Yukawa coupling, the top quark mass
and $\tan\beta$ are predicted to be large.  As shown in Figure 1,
$166$ GeV $< M_t < 192$ GeV, and $51 <\tan\beta < 63$.
 The predictions for the top quark mass and $\tan\beta$
shown in Figure 1, arise when all three third generations Yukawa couplings
are equal at the grand unification scale.  Perturbative corrections to this
equality occur when the operator $O_{23}$ has large Clebsch factors since
the third generation Yukawa couplings are equal to
$$
\lambda_{i}^2 = A^2  + (x_{i}^2 + x'_{i}^2)B^2
\eqno(6.1)
$$
at the unification scale.  This effect is numerically insignificant in most
models, because typically $x_i B/A \approx V_{cb}$. However, $V_{cb}$ actually
constrains only $(x_d - x_u)B/A$ and occasionally some $x_i$ and $x_i'$
Clebsch factors, particularly those with $i=e$, happen to be large.
Of the nine models this is most significant in model 9 where the top quark
mass prediction is decreased by $5-10$ GeV.

The nine ``22" texture models separate
into three groups according to
their predictions for $V_{cb}$:  $V_{cb} = \chi \sqrt{m_c/m_t}$ with
$\chi=  2/3$, $ 5/6, 8/9$.  These three classes of $V_{cb}$ predictions are
shown in Figure 4 along with the prediction for $\chi = 1$.   The values
of $V_{cb}$ obtained from exclusive $B$ decays using heavy quark
effective theory favors the $\chi = 2/3$ and $\chi = 5/6$ models.
Figure 4 also shows that the HRR\cite{GJ} relation
$V_{cb} = \sqrt{m_c/m_t}$ is incompatible with a framework which unifies
the third generation Yukawa couplings because it leads to unacceptably
large values of $V_{cb}$.

  A second feature which distinguishes
the nine models  is the predictions for $m_s/m_d$.
We can compare these predictions to the chiral perturbation theory
determination of light quark mass ratios.  Second order chiral
perturbation theory along with Dashen's theorem can be used to determine
the light quark mass ratios $m_s/m_d$ and $m_u/m_d$ in terms of the
pseudoscalar meson masses\cite{KM,L}:
$$
\frac{1}{Q^2} \left( \frac{m_s}{m_d} \right)^2
            + \; \left( \frac{m_u}{m_d} \right)^2 = 1
\eqno(6.2a)
$$
where
$$
     Q^2 = \frac{M_K^2}{M_\pi^2}
 \frac{M_K^2 - M_\pi^2}{M_{K^O}^2-M_{K^+}^2 +M_{\pi^+}^2 -M_{\pi^0}^2}
         \simeq 24^2
\eqno(6.2b)
$$
The theoretical uncertainties in Eq. 6.2 can be cast as an uncertainty
in $Q$.  Figures 5a, 5b and 5c compare the light quark mass ratio
predictions of the nine ``22" models.  When the inputs are restricted
to the ranges $m_b(m_b) = 4.25 \pm 0.1$ GeV, $m_c(m_c) = 1.27 \pm 0.05$ GeV and
$\alpha_s(M_Z) = 0.120 \pm 0.01$, only models 2,4,6 and 9 can yield
successful predictions with models 4,6 and 9 providing the most
successful predictions. The prediction of the light quark mass ratios gives a
preference for low values for $\alpha_s(M_Z)$ in all models.

   As discussed at the end of this section,
the $\chi = 2/3$ models do not have enough
CP violation to explain the kaon CP impurity parameter $\epsilon$.
So the requirement that these models account for the observed CP
violation eliminates models 1-4, leaving us with two remaining
candidates: 6 and 9. As an important discriminator, we show that
correlations between the
predictions for $V_{cb}$ and $m_u/m_d$ tends to disfavor model 9.
For all nine models, predictions for $m_u/m_d$ small enough to agree
well with the chiral perturbation theory determination occur
for large values of the inputs $m_b(m_b)$ and $m_c(m_c)$,
and small values of $\alpha_s(M_Z)$. Each of these leads
to a larger value of $V_{cb}$.  For example,
with model 9 $m_u/m_d < 0.6$ requires $V_{cb} > 0.052$.
This competition is shown in Figures 6 and 7 for models
9 and 6 respectively. It may be over zealous to say that model
9 is ruled out by this analysis because there are uncertainties
in these calculations at the level of a few percent, and we
have used a conservative range of inputs in this analysis.
However, the empirical value of fermion masses and mixing angles
seems to show a preference for model 6.

At the level where they can be probed experimentally,
the predictions for $m_s$ and $V_{ub}/V_{cb}$ are fairly universal
between the nine ``22" models.  For the sake of definiteness,
the model 6 predictions  are shown in Figures 8 and 9.  Both
predictions are in excellent agreement with current
determinations.  The successful prediction for the strange quark mass
reflects the Georgi-Jarlskog factor of three, which is incorporated in the
operator $O_{22}$.

The final prediction for these models is for $\epsilon$, the measure of
$CP$ impurity in $K_L$ and $K_S$. The dominant contribution to $\epsilon$ comes
from the standard model box diagram amplitude, which may be written as a
function of CKM elements and quark masses multiplied by a QCD matrix element
$B_K$ \cite{BK}. Using this result, together with
the measured value of $\epsilon$ and our predictions for
quark masses and CKM parameters, we can give a
prediction for the matrix element $B_K$. In the theoretical formula for
$\epsilon$ \cite{BH} we use input quantities renormalized
at $\mu$ such that $\alpha_s(\mu) = 1$, hence the predicted quantity is
$\hat{B_K}$. Figure 10
shows the range of predictions for $\hat{B_K}$ for the nine ``22" models.
Five of the nine models predict values of $\hat{B_K}$ that agree well with
the lattice determination of
$\hat{B}_K = 0.72 \pm 0.06$ \cite{S}, where the
error includes only the uncertainty due to the continuum extrapolation.
Other calculational approximations, such as the quenched
approximation, lead to additional uncertainties.
The $\chi  = 2/3 $ models give insufficient $CP$ violation,
and can be excluded if we demand that the observed CP violation in
the kaon system arise from the $CP$ violating phase in the CKM matrix.

Predictions for $\sin 2\alpha$ and $\sin 2\beta$ provide a
complimentary method of distinguishing and testing the nine ``22" models.
Most importantly, measurements of these angles from $CP$ violating
asymmetries in B decays  provide a potentially more stringent
test of these models.  Figures 11a, 11b, and 11c display the
$\sin 2\alpha$ vs. $\sin 2\beta$ predictions of models 1-9. Also shown is the
expected size of the experimental error bar for an integrated
luminosity of $10^{41}$ cm$^{-2}$ at an asymmetric B factory.
Such measurements could exclude our entire scheme.
They could also distinguish between some
(e.g. 5 and 6) but not all (e.g. 7 and 8) models.

Finally, we provide three examples of particular predictions
to demonstrate the degree of simultaneous success each class of model
can achieve.

\begin{center}
\begin{large}
\indent {\bf Table 3: Particular Predictions for Model 4
with $\alpha_s(M_Z) = 0.110$}
\vskip 20pt
\begin{tabular}{|c|c|c|c|}
\hline
  Input Quantity & Input Value & Predicted Quantity & Predicted Value\cr
\hline
  $m_b(m_b)$ & $4.35 $ GeV & $M_t$ & $166$ GeV \cr
  $m_\tau(m_\tau)$ &$1.777 $ GeV & $\tan\beta$ & $51 $  \cr
\hline
  $m_c(m_c)$ &$1.32$ GeV & $V_{cb}$ & $.043 $ \cr
\hline
  $m_\mu$ &$105.6 $ MeV   & $V_{ub}/V_{cb}$ & $.046 $ \cr
  $m_e $   &$0.511$ MeV  &  $m_s(1 GeV)$ & $147 $ MeV \cr
  $V_{us}$       &$0.221 $    & $\hat{B_K} $ & $1.1$  \cr
                 &            & $m_u / m_d$  & $0.41$  \cr
                 &            & $m_s / m_d$  & $22.$  \cr
\hline
\end{tabular}
\end{large}
\end{center}
In addition to these predictions, the set of inputs in
Table 3 predicts:

$\sin 2\alpha = -.32$, $\sin 2\beta = .39$,
$\sin 2\gamma = .66$, and $J = 1.75\times 10^{-5}$.

\begin{center}
\begin{large}
\indent {\bf Table 4: Particular Predictions for Model 6
with $\alpha_s(M_Z) = 0.115$}
\vskip 20pt
\begin{tabular}{|c|c|c|c|}
\hline
  Input Quantity & Input Value & Predicted Quantity & Predicted Value\cr
\hline
  $m_b(m_b)$ & $4.35 $ GeV & $M_t$ & $176$ GeV \cr
  $m_\tau(m_\tau)$ &$1.777 $ GeV & $\tan\beta$ & $55 $  \cr
\hline
  $m_c(m_c)$ &$1.22$ GeV & $V_{cb}$ & $.048 $ \cr
\hline
  $m_\mu$ &$105.6 $ MeV   & $V_{ub}/V_{cb}$ & $.059 $ \cr
  $m_e $   &$0.511$ MeV  &  $m_s(1GeV)$ & $172 $ MeV \cr
  $V_{us}$       &$0.221 $    & $\hat{B_K} $ & $0.64$  \cr
                 &            & $m_u / m_d$  & $0.64$  \cr
                 &            & $m_s / m_d$  & $24.$  \cr
\hline
\end{tabular}
\end{large}
\end{center}
In addition to these predictions, the set of inputs in
Table 4 predicts:

$\sin 2\alpha = -.46$, $\sin 2\beta = .49$,
$\sin 2\gamma = .84$, and $J = 2.6\times 10^{-5}$.

\begin{center}
\begin{large}
\indent {\bf Table 5: Particular Predictions for Model 9
with $\alpha_s(M_Z) = 0.120$}
\vskip 20pt
\begin{tabular}{|c|c|c|c|}
\hline
  Input Quantity & Input Value & Predicted Quantity & Predicted Value\cr
\hline
  $m_b(m_b)$ & $4.35 $ GeV & $M_t$ & $180$ GeV \cr
  $m_\tau(m_\tau)$ &$1.777 $ GeV & $\tan\beta$ & $58 $  \cr
\hline
  $m_c(m_c)$ &$1.27$ GeV & $V_{cb}$ & $.050 $ \cr
\hline
  $m_\mu$ &$105.6 $ MeV   & $V_{ub}/V_{cb}$ & $.071 $ \cr
  $m_e $   &$0.511$ MeV  &  $m_s(1GeV)$ & $172 $ MeV \cr
  $V_{us}$       &$0.221 $    & $\hat{B_K} $ & $0.43$  \cr
                 &            & $m_u / m_d$  & $0.75$  \cr
                 &            & $m_s / m_d$  & $23.$  \cr
\hline
\end{tabular}
\end{large}
\end{center}
In addition to these predictions, the set of inputs in
Table 5 predicts:

$\sin 2\alpha = -.14$, $\sin 2\beta = .59$,
$\sin 2\gamma = .70$, and $J = 3.6\times 10^{-5}$.

\vskip 9pt
\noindent{\bf (7) THE ``23'' TEXTURE}
\vskip 9pt
In this section we study the second GUT scale Yukawa texture
which could arise from an SO(10) flavor sector with just four operators.
The four operators of this ``23" texture are shown in equation (3.2), and lead
to Yukawa matrices which
can be written in a form analogous to (5.1):

$$ \lambda_i = \pmatrix{0&z_i^{\prime} C&0\cr
    z_i C&0&x_i^{\prime}exp(i \phi_i^{\prime})B\cr
    0&x_i exp(i \phi_i) B&A\cr}\hfil \eqno(7.1)$$
where $i=u,d,e$.
$$x_i^{(\prime)}e^{i\phi_i^{(\prime)}} = x_{ia}^{(\prime)} +
\gamma e^{i\phi} x_{ib}^{(\prime)}
\eqno(7.2)$$
is the complex linear combination of Clebsch factors arising from the
operator $O_{23} = O^a_{23} + \gamma e^{i\phi} O^b_{23}$.
(Strictly speaking, we should distinguish between the
$\gamma$ which appears in the linear combination of operators and
the $\gamma$ which appears in the linear combination of Clebsches
contributing to the Yukawas entries.  These $\gamma$s are actually related
by a real multiplicative constant, because the Higgs vacuum expectation values
which contribute to $O_{23}^{(a,b)}$ may be differently normalized.  We absorb
this factor into the definition of $\gamma$ in (7.2).)
For any given model,  the values of $x_i^{(\prime)}$ (and $z_i^{(\prime)}$)
can be obtained from Table 1.
By making phase rotations on the left and right matter multiplets,
this form for $\lambda_i$
can be brought into that of (5.1) ($\phi_i,\phi^{\prime}_i = 0$)
with $y_i = 0$:
$$ \lambda_i \to
\pmatrix{e^{i\phi_i} & 0 & 0\cr 0 & e^{-i\phi_i^{\prime}} &
0 \cr 0 & 0 & 1\cr}
                \lambda_i
\pmatrix{e^{i\phi_i^{\prime}} & 0 & 0\cr 0 & e^{-i\phi_i} &
0 \cr 0 & 0 & 1\cr}.
$$
Approximate diagonal forms for these matrices at the GUT scale
can then be obtained by essentially the same sequence
of transformations as was used to diagonalize the matrices (5.1):
\hfil \break i) In the heavy $2\times2$ sector rotate the
left-handed fermions by angles
$\alpha_i = x_i B/A$.
and the right-handed ones by $\alpha_i^{\prime} = x_i^{\prime} B/A$.
\hfil \break
ii) Write the resulting $22$ entry as $y_i E_i$.  \hfil \break
ii) Diagonalize the light $2\times2$ sector by rotations
$\beta_i = z_iC/y_iE_i$ and $\beta_i^{\prime} = z_i^{\prime}C/y_iE_i$
on the left and right-handed fermions respectively.

Let ${\bar \theta_3} = -\alpha_u$, ${\bar \theta_4} = -\alpha_d$
$\theta_2 = -\beta_u$, $\theta_1 = -\beta_d$, then the transformations
in the left-handed up and down sector are:
$$V_u = \pmatrix{c_2& s_2 & 0\cr -s_2&c_2&0\cr 0 & 0 & 1\cr}
        \pmatrix{1&0&0\cr 0& {\bar{c_3}}&{\bar{s_3}}\cr
                         0&-{\bar{s_3}}&{\bar{c_3}}\cr}
\pmatrix{e^{-i\phi_u^{\prime}} & 0 & 0\cr 0 & e^{i\phi_u} &
0 \cr 0 & 0 & 1\cr}.
\eqno(7.3a)
$$
$$V_d = \pmatrix{c_1& -s_1 & 0\cr s_1&c_1&0\cr 0 & 0 & 1\cr}
        \pmatrix{1&0&0\cr 0& {\bar{c_4}}&{\bar{s_4}}\cr
                         0&-{\bar{s_4}}&{\bar{c_4}}\cr}
\pmatrix{e^{-i\phi_d^{\prime}} & 0 & 0\cr 0 & e^{i\phi_d} &
0 \cr 0 & 0 & 1\cr}. \eqno(7.3b)
$$
The resulting KM matrix, $V = V_u V_d^{\dagger}$, is not of the form  (4.3)
for general choices of $\phi_i, \phi^{\prime}_i$ (although one more phase
redefinition can bring it into this form).

As discussed above and proved in appendix 3,
once the 33 operator has been chosen
the choice of the 12 operator is unique.
It is only the choice of 23 operators which remains to determine the model.
This  selection is constrained primarily by the crucial relation
(5.4e),
$$
V^2_{cb_G} \simeq \left| {\lambda_c\over \lambda_t}\right|_G  \ll
\left| {\lambda_\mu\over \lambda_\tau}\right|_G\simeq 3
\left| {\lambda_s\over\lambda_b}\right|_G \eqno(5.4e)
$$
which indicates a division into two small parameters and two large
ones.  In the case of the  \lq\lq22 texture\rq\rq, the small
parameters were obtained from the 23 operators, and the large
ones from the 22 operators.  Here, no such separation is possible;
all four parameters are generated by the same linear combination
of operators.   The hierarchy must therefore result from
a set of Clebsches which allow
cancellations between the contributions from
$O_{23}^a$ and $O_{23}^b$ to both $V_{cb}$ and $\lambda_c/\lambda_t$,
for similar values of $\gamma exp[i\phi]$.
This cancellation implies that the hierarchy of equation (5.4e) is unnatural,
and it is for this reason that we prefer the ``22" models to the ``23" models.
%The constraints of this situation have the attractive feature that
%$\gamma$ and  $\phi$ are not truly independent parameters in this class
%of models.  For each model there is an additional, parameter-independent
%\lq\lq{prediction}\rq\rq ~lurking in the enforced cancellations of the 23
%operators.

   Unfortunately, the complex nature of $O_{23}$
in the 23 texture also makes these models considerably less amenable
to analytic treatment than were the 22 models.  The reason is clear.
Whereas in the 22 models (equation 5.1) the complex phase $\phi$
which appeared in the 22 entry of $\lambda_i$ was the same for
$i=u,d,e$, in the 23 models the one input phase $\phi$ appears
as six distinct phases $\phi_i, \phi^{\prime}_i$ in the Yukawa
matrices.  The relation of these $\phi^{(\prime)}_i$ to $\phi$
is Clebsch-dependent and hence depends on the particular model.

   We have therefore undertaken a numerical search of all
possible dimension-five and six operators $O_{23}^{a,b}$ satisfying
our dynamical hypotheses.
(There are  approximately twenty-thousand such operator pairs.)
If any significant fraction of these models reproduced the
available experimental data, then the probability of being able
to make useful predictions from this approach would be small;
for any realistically achievable experimental error-bars,
the predictions would likely be dense   in the experimental plane.
Fortunately, it is very unlikely that any individual model will
work.

 The search was performed by first considering the GUT-scale predictions
for the quark and charged-lepton
mass ratios and mixing angles, which
(to leading order in the small quantities $(B/A)$ and $(C/A)$) are:
$$\eqalignno{
{\lambda_s/\lambda_b\over\lambda_\mu/\lambda_\tau}
= {\lambda_s\over\lambda_\mu}
     &= {\vert x_dx_d^{\prime}\vert\over\vert x_ex_e^{\prime}\vert}
&(7.4a)\cr
{\lambda_c/\lambda_t\over\lambda_s/\lambda_b}
= {\lambda_c \over \lambda_s}
     &= {\vert x_ux_u^{\prime}\vert\over \vert x_dx_d^{\prime}\vert}
&(7.4b)\cr
{\Vcb^2\over\lambda_c/\lambda_t} \phantom{= {\lambda_c/\lambda_s}}
     &= {\vert x_d e^{i\phi_d}-x_u e^{i\phi_u}\vert^2
        \over\vert x_ux_u^{\prime}\vert}
&(7.4c)\cr
{\sinc\over{\sqrt {\lambda_e/\lambda_\mu}}} \phantom{= {\lambda_c/\lambda_s}}
     &= \biggl\vert{1\over x_dx_d^{\prime}e^{i(\phi_d+\phi_d^{\prime})}} +
        {1/27\over x_u x_u^{\prime}e^{i(\phi_u+\phi_u^{\prime})}}\biggr\vert
                                \bigl\vert x_ex_e^{\prime}\bigr\vert
&(7.4d)\cr
}$$
(The predictions for the first generation masses being fixed by the 12
operator.)
Here all Yukawa couplings $\lambda_a$ are evaluated at the GUT scale.

Although these expressions involve ratios of masses at the GUT scale,
we understand their running well enough to be able to make some
definitive statements.
In particular $(\lambda_e/\lambda_\mu)=(\me/\mmu)$   and $\sinc$
essentially do not run from the GUT scale down to low energies,
so that $(\sinc/{\sqrt{\me/\mmu}})_G \simeq 0.015$,
while as discussed above (equation (5.4e)) $(\lambda_s/\lambda_\mu)_G = 1/3$ in
any theory where $\lambda_t=\lambda_b=\lambda_{\tau}$ at $\MGUT$.
Since
$x_i^{(\prime)}e^{i\phi_i^{(\prime)}}
 = x_{ia}^{(\prime)} + \gamma e^{i\phi} x_{ib}^{(\prime)}$,
equations (7.2a,d) determine $\gamma e^{i\phi}$.  Unfortunately
they cannot be solved analytically for arbitrary values of
$x_{ia}^{(\prime)}$ and $x_{ib}^{(\prime)}$.
Therefore for each possible choice of 23-operator pair we
solved equations (7.2a,d) numerically
to find the  allowed value(s) of $\gamma$ and $\phi$,  for
a small range of input values about the central values given above.
This range allows for experimental uncertainties,
for any small amount of renormalization, and for the difference between
the exact and leading order eigenvalues of the $\lambda^i$ and
the rotations needed to diagonalize them.
(The latter being particularly crucial for $\sinc$.)
For each case we
either obtained values for $\gamma$ and $\phi$ or discovered that
the equations have no solution. For those models for which
values of $\gamma$ and $\phi$ were  obtained,
equations (7.2b,c) were used to predict the
GUT-scale values of $(\lambda_c/\lambda_s)$
and $\Vcb/{\sqrt{(\lambda_c/\lambda_t)}}\equiv\chi$.
As shown above (equations 5.4 and 5.11),
the allowed ranges for these quantities are
$0.55 < \chi < 0.92$ and
$0.085 \lsim (\lambda_c/\lambda_s)_G \lsim 0.19$.

Demanding
that $(\lambda_c/\lambda_s)_G$
and $\chi$ lie within their allowed ranges,
we were able to reduce the number of possible 23 models to less than one
hundred.  These we explored individually by running them
down from the GUT scale to low energy using 2-loop renormalization
group equations and appropriate threshold effects.  The values of
$A$ and  $\tanb$ were determined (as described in section 4 above)
by fixing the values of $(\mb/\mtau)$ and $\mtau$.
The remaining four parameters of the models --
$B$, $C$, $\gamma$, and $\phi$  --  were used to fit
$\mc$, $\mmu$, $\me$ and $\sinc$.
Predictions were obtained for
the eight remaining physical quantities in the quark-lepton mass sector:
$\mtop$, $\ms$,
$\stod$, $\utod$, $\Vcb$, $\Vub/\Vcb$, $\tanb$ and $B_K$ (or the $J$).
Most of the models are found to fail for one reason or another.
For only four choices of
23 operators can the six inputs quantities be self-consistently
reproduced as well as giving $V_{cb} = .043 \pm .007$ and  values of
$m_d/m_s$ \versus $m_u/m_d$ consistent with experiment:
$$\eqalignno{
A)\; 16_2 45_1 10 (1/45_1) 16_3
&+ \gamma e^{i\phi} 16_2 10 45_1 45_{24} 16_3
&(7.5a)\cr
B)\; 16_2 45_1 45_1 10 16_3
&+ \gamma e^{i\phi} 16_2 45_{24} 10 45_{24} 16_3
&(7.5b)\cr
C)\; 16_2 45_1 45_{24} 10 16_3
&+ \gamma e^{i\phi} 16_2 45_1 10 (45_{B-L}/45_{24}) 16_3
&(7.5c)\cr
D)\; 16_2 45_1 45_1 10 16_3
&+ \gamma e^{i\phi} 16_2 45_{24} 10 (1/45_1) 16_3
&(7.5d) \cr
}$$

Of these four models, (D) can reasonably be excluded because
it predicts $V_{ub}/V_{cb }\leq .035$, considerably smaller than the
accepted experimental value.  $V_{ub}/V_{cb }$ cannot be increased without
increasing
$V_{cb}$ above $0.053$.  The other three are consistent with all available
experimental data against which we have compared them.

The models (A), (B) and (C) make 7 successful predictions
for flavor parameters.
We present  in figures 12, 13, and 14 the
predictions for $m_t(m_t)$,  $\tan \beta$,
$m_s(1\GeV)$, $\Vcb$ and  $\Vub/\Vcb$ \versus $\alpha_3(M_Z)$, and
$\stod$ vs $\utod$,
for models (A), (B) and (C) respectively.
Several observations are in order.
First, there is a definite lack of universality
in the predictions for $\mtop$ and $\tanb$ \versus $\althreemZ$.
Model (A) in particular shows
notably different behaviour than (B) and (C), with a wider range of
permissible values for both $m_t$ and $\tanb$,
and consequently for $\althreemZ$.
In fact, if future experiments support larger values of
$\althreemZ$ ($\geq 0.12$), then only model (A) would remain viable,
with a definite prediction of large values for both $m_t$ and $\tanb$.

In all three ``23" models the top quark can be lighter than in the ``22"
models. This is due to the modification of the third generation Yukawa coupling
eigenvalues by the 23 entries in the Yukawa matrices. In the ``22" models
these 23 Yukawa entries are typically small, because they are generating the
small quantities $V_{cb}^2$ and $\lambda_c / \lambda_t$. Hence the correction
which they provide to the third generation Yukawa eigenvalues is small (with
the notable exception of model 9 where $x_e$ is anomalously large). In the
``23" models the situation is considerably different: the 23 Yukawa entries
must be much larger so that they can account for the larger quantities
$\lambda_\mu / \lambda_\tau$ and $\lambda_s / \lambda_b$. The smallness of
$V_{cb}^2$ and $\lambda_c / \lambda_t$ is due to an accidental cancellation, as
we have stressed. Thus in the ``23" models the larger 23 Yukawa entries give
considerable corrections to the third generation Yukawa eigenvalues, allowing
top quark masses as low as 145 GeV.

All three of the ``23" models
give remarkably good agreement with the values of $m_s/m_d$ \versus
$m_u/m_d$ derived from experiment.  These are unlikely to
be useful in distinguishing between these models.

Tests of these models will come from more accurate measurements of
CP violation, $\Vcb$ and $\Vub/\Vcb$. The predictions for these quantities are
correlated since in any given model they must arise from the same set of
inputs, in particular, the same value of $\alpha_3(M_Z)$.
Using the experimental value for the kaon CP impurity parameter $\epsilon$,
we can predict the QCD matrix element $\hat{B}_K$, which is shown in
figure 15 for each of the three models.
All are consistent with the lattice result
$\hat{B}_K = 0.72 \pm 0.06$ \cite{S}, where the
error includes only the uncertainty due to the continuum extrapolation.
Other calculational approximations, such as the quenched
approximation, lead to additional uncertainties.
Notice that the prediction of $\hat{B}_K$ is less precise in the ``23" models
th
an in
the ``22" models. This is presumably a reflection of the fact that the ``23"
models involve a degree of parameter tuning and have predictions which
therefore depend sensitively on the inputs.
In figures 16, 17 and 18 we show respectively
the  correlation of $\hat{B}_K$ with $\Vcb$ , of
$\hat{B}_K$ with $\Vub/\Vcb$ and of $\Vub/
\Vcb$ with $\Vcb$.
The very tight relationship between $\hat{B}_K$ and
$\Vcb$, particularly for models B and C, should allow these models to be
tested , though not distinguished from each other.
If $\hat{B}_K \lsim 0.7$, then all three models can be distinguished on the
basis of their predictions for $\Vub/\Vcb$. Model B can be distinguished in any
cases by its prediction of $\Vub/\Vcb \simeq 0.08$,  the others giving $\Vub/
\Vcb \lsim 0.07$ for $\hat{B}_K \leq 1.5$.

In figure 19 we plot $B_{B_d}F_B^2/\GeV^2$, which for fixed $x_d$ is a
measure
of $B_d^0{\bar B}_d^0$ mixing (cf equation (5.44)).
$F_B$ is normalized such that $F_K = 165$ MeV.
The most important consequence is that
for large values of $\althreemZ$, model A favors larger values of $B_{Bd}F_B^2/
\GeV^2$.

Finally we consider CP violation in the decays of neutral B mesons.
Figure 20 shows the predictions for $\sin2\alpha$ \versus $\sin2\beta$.
Here, model B and C can be separated from each other,
since model B predicts larger values of $\sin 2 \beta$ than model C. This is
largely a reflection of model B having larger $V_{ub}/V_{cb}$. In fact much of
the model C region with very low $\sin 2 \beta$ is excluded because
$V_{ub}/V_{cb}$ is too small.

As a demonstration that successful predictions can be obtained
simultaneously we provide an example of a particular set of predictions.

\begin{center}
\begin{large}
\indent {\bf Table 6: Examples of Predictions for Models
A,B and C}
\vskip 20pt
\begin{tabular}{|c|c|c|c|c|c|c|}
\hline
\phantom {quantity} & Model A  & Model B  & Model C  \cr
\hline
  $\alpha_1(M_{GUT})$ &0.0411&0.0412&0.0412 \cr
  $\alpha_2(M_{GUT})$ &0.0411&0.0412&0.0412 \cr
  $\alpha_3(M_{GUT})$ &0.0400&0.0396&0.0397 \cr
  $M_{GUT}/\GeV$ &$1.65\times 10^{16}$&$1.65\times 10^{16}$&$1.65\times
10^{16}$ \cr
  $M_{SUSY}/\GeV$ &$200$&$200$&$200$ \cr
  $\alpha_3(M_Z)$ &0.110&0.106&0.107 \cr
  $\sin^2\theta_W$ &.2325&0.2325&0.2325 \cr
\hline
  $A$ &0.80&0.55&0.575 \cr
  $B/A$ &0.02655&0.04544&0.04202 \cr
  $C/A$ &$1.223\times 10^{-4}$&$1.485\times 10^{-4}$&$1.865\times 10^{-4}$ \cr
  $\gamma$ &0.931&0.139&0.734 \cr
  $\phi$ &3.946&-1.782&0.5125 \cr
\hline
  $M_t/\GeV$ &$173$&159&$162 $ \cr
  $m_b(m_b)/\GeV$ &$4.16$&4.26& $4.32 $ \cr
  $m_\tau(m_\tau)/\GeV$ &1.7841 &1.7841 &1.7841 \cr
  ${\rm tan}\beta $   &55.16&50.01&49.75\cr
\hline
\end{tabular}
\newpage
\vskip 20pt
\begin{tabular}{|c|c|c|c|c|c|c|}
\hline
\phantom {quantity} & Model A  & Model B  & Model C  \cr
\hline
  $m_c(m_c)/\GeV$ &1.29&1.29&1.27 \cr
  $m_s/\MeV$ &140&137&141 \cr
  $m_\mu/\MeV$ &105.658&105.658&105.658 \cr
  $V_{cb}$ &0.044&0.046& 0.044 \cr
\hline
  $m_s / m_d$ &19.7&21.8& 22.9 \cr
  $m_u / m_d$ &0.511&0.451& 0.441 \cr
  $m_e/\MeV $   &0.511&0.511& 0.511  \cr
  $V_{ub}/V_{cb}$ &0.046&0.081& 0.054 \cr
  $V_{us}$ &0.2210&0.2202& 0.2214 \cr
\hline
  $B_K$ &0.92&0.65& 0.86 \cr
  $\sin 2\alpha$ &0.32&0.98& 0.48 \cr
  $\sin 2\beta$ &0.39&0.53&  0.45 \cr
  $\sin 2\gamma$ &0.08&-0.74& -0.03 \cr
  $B_{Bd}F_B^2 /\GeV^2$ &0.033&0.045& 0.037 \cr
\hline
\end{tabular}
\end{large}
\end{center}

In this section we have taken all superpartners degenerate at 200 GeV.
These predictions of third generation mixing and of CP violation
are not greatly modified by threshold effects at the weak
and/or SUSY scales.

In conclusion, three models of the 23 texture reproduce all the observed
masses and mixings of the quarks and leptons.
The models give remarkably good agreement for
the values of $\stod$ \versus $\utod$.  They are best tested by the combination
of the predictions for $\Vcb$, $\Vub/\Vcb$, and $\hat{B}_K$,
and  of $\sin2\beta$ \versus $\sin2\alpha$.

\vskip 9pt
\noindent{\bf (8) CONCLUSIONS}

In this paper we have explored the consequences of an effective
SO(10) grand unified theory with family symmetries which yield very
simple and economical flavor structures predicting 7 of the 13 flavor
parameters of the standard model.
In section 3 we gave an explicit list
of eight assumptions which the theory must satisfy in order that this maximal
number of flavor predictions result. Some of these assumptions we believe to be
mild and well motivated: they yield the simplest picture in which the weak
mixing angle is a significant prediction. Other assumptions are stronger
and could easily be violated: for
example, it may be that the pattern of family symmetries allows further
important flavor operators, or
the two Higgs doublets may not lie completely in a single 10 dimensional
representation. As each assumption is relaxed the whole picture is not
destroyed, rather additional free parameters must be added which successively
reduce the predictivity of the theory. We have no compelling reason for why
nature should choose the most predictive case; nevertheless, it is appealing to
suppose that nature is simple and
we find it quite
striking that such a predictive possibility is allowed by experiment. We are
excited by the prospect that a combination of experiments, each
designed to make accurate
measurements of parameters of the standard model, could reveal a very simple
group theoretic structure underlying the masses and mixings of the quarks and
leptons.

In this paper we have discovered and elucidated the predictions of two classes
of SO(10) theories which have seven flavor predictions. Every model we have
constructed has just a single renormalizable Yukawa coupling which is
responsible for $m_t, m_b$ and $m_\tau$ \cite{ALS,HRS} and, in most models,
this results in a heavy top quark: $M_t = 180 \pm 15$ GeV.
We find this picture of third
generation Yukawa coupling unification to be rather elegant; however in
Appendix 4 we warn the reader that it is quite possible to retain the
elegance while losing the top mass prediction!

Each of our models has three
further operators in the flavor sector, and the models fall into two classes
according to whether the resulting texture is of the ``22" or ``23" type (shown
in eq. (3.1) and (3.2) ).  These additional operators are non-renormalizable,
and have dimension and flavor structure allowing the observed hierarchy of
quark and lepton masses and mixings to emerge as
a consequence of the small ratio of the grand to Planck mass scales $M_G/ M_P$.
The flavor structure of the operators is dictated in a very straightforward way
from the observed pattern of nearest neighbour mixing of the CKM matrix:
there must be at least one operator which
mixes the heaviest two generations $O_{23}$ and at least one which mixes the
lightest two $O_{12}$. Furthermore, in appendix 2 we prove that the fourth
operator must be
$O_{22}$ or  $O_{23}'$, leading to the two different textures.
A very general argument in appendix 1 shows that the operator  $O_{12}$ is
unique, and therefore common to both textures. For models with the ``22"
texture
the operator $O_{22}$ must also lead to unique Yukawa
coupling relations. Hence the
multiplicity of models which we have discovered, 9 for the ``22" texture and 3
for the ``23" texture, is a reflection of the many differing operators for the
23 entry.

How significant is the finding of this paper: that the wealth of experimental
data on the masses, mixings and CP violation of the quarks and leptons can be
described in terms of just four SO(10) invariant operators? The operator search
apparently ranges over a vast number of possible operators. If models can
typically be found for any values of the masses and mixings then our result is
not particularly surprising or significant. This is not the case; the argument
is as follows. Models with the minimal number of flavor operators must be of
the ``22" or ``23" texture. The number of possible operators for $O_{33}$,
$O_{23}$ ,$O_{22}$ and $O_{12}$ are 2, 152, 108 and $\approx 3.10^4$
respectively. Only one of the $O_{33}$ operators can accommodate
the observed value for $m_b/m_\tau$, certainly it would not be possible to
accommodate any value. Of the $\approx 3.10^4$ possible $O_{12}$ operators only
one is consistent with $60 MeV < m_s < 360 MeV,  \; 0.2<m_u/m_d<1.5$ and the
observed Cabibbo angle. If $m_u/m_d$ had been, say, 3, our whole program would
have collapsed. The uniqueness of $O_{12}$ we believe to be one of our most
interesting results. Next consider the operator $O_{22}$ of the ``22" texture.
It generates the ``large" parameters $m_s/m_b$ and $m_\mu/m_\tau$, but is not
responsible for the ``small" parameters $m_c/m_t$ and $V_{cb}^2$, and hence
must give the Clebsch ratios close to the uniquely successful ones
of eq. (5.5): $y_u : y_d : y_e = 0:1:3$.
Suppose that the experimental value of the muon mass had turned out to be
twice the usual number, then we would be seeking Clebsch ratios
close to $y_u : y_d : y_e = 0:1:6$. There are no
operators which give Clebsches close to this. If the muon mass were doubled (or
trebled) our program would fail. The success and uniqueness of the operators
$O_{33}$, $O_{22}$ and $O_{12}$ is significant and by no means guaranteed.
The situation with the operator  $O_{23}$ is
somewhat different. It is not unique: there
are 9 possible 23 operators in the ``22"
textures and three such possible pairs in
the ``23" texture. This operator generates $V_{cb}$. If $V_{cb}$ were doubled
we could still find acceptable  $O_{23}$ operators for the ``22" texture.
However, if $V_{cb}$ were halved there would be no acceptable operator and our
program would again fail. The value of $V_{cb}$ obtained in models 1-4 is the
lowest which we can obtain, and these models tend not to have sufficient CP
violation. A more accurate experimental value of $V_{cb}$ is of very great
interest: a reduced error bar will tell us which, if any, $O_{23}$ operator is
correct. A low experimental value for $V_{cb}$ would be sufficient to exclude
models 5-9, which give a range $0.045 < V_{cb} < 0.055$.

While models of both textures can be found which agree well with data, we have
a theoretical bias towards the models with ``22" texture, as those with the
``23" texture involve a modest fine-tuning, between the coefficients of
$O_{23}$ and $O'_{23}$,
in order to understand the smallness of $V_{cb}$.
The predictions for the models with ``22" texture are given with approximate
analytic equations in section 5 and from a numerical calculation in figures 1
to 11. All possible models have been found which satisfy the search criteria of
section 3a.
Of the nine ``22" models,
four (models 1-4) have difficulty in yielding
sufficient CP violation in the neutral K
system. The remaining 5 models give $0.045 < V_{cb} < 0.055$.
The predictions for the three models with ``23" texture are discussed
in section 7 and  numerical results are shown in figures 12 to 20.

It will not be an easy experimental task to distinguish between the models.
Nevertheless we believe that a real test of these models is possible.
Each model has 7 flavor predictions, with theoretical uncertainties of about
10\%. Some theoretical uncertainties at this level are expected from physics at
the grand unification scale and hence, without a more detailed theory,
probably cannot be removed. It is therefore crucial that these models be tested
by the combination of all the predictions. The first stage will involve testing
whether any model successfully accommodates improved data on $M_t, V_{cb}$ and
$V_{ub}/V_{cb}$ with an improved value of the strong coupling, which must also
be consistent with the $m_s/m_d$ versus $m_u/m_d$ plot. The second stage will
be
to test the predictions of the models for the CP violating angles $\sin 2
\alpha, \sin 2 \beta$ and $\sin 2 \gamma$ in neutral B meson decay, and to test
the predictions for $B^0 \bar{B}^0$ mixing, for both $B_d$ and $B_s$.

We have not considered neutrino masses in this paper. Dirac neutrino mass
matrices, which mix the left-handed electroweak doublet neutrinos with the
right-handed singlet neutrinos, are already fixed by our analysis. However, in
order to obtain Majorana masses for the right-handed neutrinos (necessary for a
see-saw mechanism) additional operators must be considered.

The predictions of this paper are the consequence of two types of symmetries.
The SO(10) grand unified symmetry
allows us to view quarks and leptons of a given family as different aspects of
a single object, and hence relates the $ij$ entries of the Yukawa coupling
matrices ${\bf U}, {\bf D}$ and ${\bf E}$. Unfortunately this elegant grand
unified symmetry is not sufficient to yield the predictions. In addition most
of the SO(10) operators which could contribute to these Yukawa couplings should
be absent. This can be accomplished by using family symmetries, in the theory
defined at $M_P$, as shown by a specific example in appendix 1.
Indeed, if family symmetries are present one {\em expects} the
majority of such operators to typically
be absent. For example, consider a U(1) symmetry
under which the heaviest generation and the Higgs doublets are neutral, but the
lighter two generations have any positive charges. The only allowed
renormalizable Yukawa interaction is that for the heaviest generation. All
other masses and mixings must be small effects induced by higher dimension
operators. The challenge is to understand why the family symmetries should be
those which are needed to yield the pattern of non-renormalizable operators
which we have found to be singled out by experiment.

\vskip 9pt
\noindent{\bf Acknowledgements.}
\vskip 9pt
GWA thanks Uri Sarid for confirming some numerical data
and the Aspen Center for Physics.
LJH thanks Riccardo Rattazzi, Graham Ross and Uri Sarid for many valuable
conversations. LJH and SR thank Riccardo Barbieri for many illuminating
thoughts on the material of appendix 1.
GDS thanks ITP at Santa Barbara and the Aspen Center for Physics
where part of this work was done.

\vskip 9pt
\noindent{\bf Appendix 1}
\vskip 9pt

SO(10) grand unification provides a very powerful reduction in the number of
free Yukawa parameters: the elements of {\bf D} and  {\bf E} can be fixed in
terms of those of {\bf U}. However this parameter reduction by itself is not
sufficient to yield predictions:  {\bf U} is in general a complex $3
\times 3$ matrix with 18 parameters. Underlying the search in this paper, for
the SO(10) flavor sector with the minimal number of operators, is the belief
that such a simple picture will emerge from a set of family symmetries. There
must be some set of symmetries which distinguishes between the three families
and which explains why the majority of both the renormalizable and
non-renormalizable superpotential interactions are absent at the GUT scale.

This is very straightforward at the renormalizable level. A U(1) symmetry with
zero charges for the $16_3$ and $10$, and positive charges for the $16_1$ and
$16_2$, immediately leads to a single renormalizable Yukawa interaction:
$16_3 \; 10 \; 16_3$. This illustrates that it is not difficult to arrange for
most couplings to vanish, giving sparse matrices of couplings. In this appendix
we extend this idea to the non-renormalizable level and give an explicit
realization of family symmetries which leads to one of our models.
This involves going beyond the effective GUT theory and looking at how the
non-renormalizable operators are obtained
by integrating out heavy states at the scale $v_{10}$
or $M_P$. In addition to providing an understanding of the origin of the flavor
structure, this analysis shows that there are corrections to these effective
operators, and we estimate the size of such corrections.

In figure 21, we show the tree diagrams which lead to one of the
``22" texture models (model 9) with $\chi = 8/9$ discussed in section 4.  The
intermediate fermions obtain large masses (of order $v_{10}$ in B and C, and
even higher in D) and
mix with the light states by Higgs VEVs of order $M_G$.  We also display the
possible global $U(1)$ family quantum number (in the upper right) which
prohibits additional terms in the mass matrix. These diagrams are to be
understood as a perturbative technique for integrating out heavy fields and
obtaining an approximate mass matrix for the light states. A more detailed
analysis of such theories will be given in a future paper, which will provide a
complete theory including the SO(10) gauge symmetry breaking sector \cite{BHR}.

To leading order in the small ratios $M_G/v_{10}$ or $v_{10}/M_P$,  the mass
matrices for the light fermions are given in terms of the four operators:

$$\eqalignno{
O_{33}=&16_3 \ 10 \ 16_3\cr
O_{23}=&16_2 \ 10 \ {45_{B-L}^2 \over 45_1^2}\ 16_3\cr
O_{22}=&16_2 \ 10 \ {45_{B-L} \ S \over 45_1^2}\ 16_2\cr
O_{12}=& 16_1 \left({45_1\over M}\right)^3 \ 10 \left({45_1\over M}\right)^3 \
16_2 & (A1.1) \cr} $$
where $S$ is an  $SO(10)$ singlet fields with VEVs of order $M_G$. Note that it
is necessary to introduce the U(1) in the fundamental theory at $M_P$ since it
is only at this level that the ordering of operators can be constrained. For
example, it is only at the renormalizable level that symmetries can guarantee
that $O_{23}$ in equation (A1.1) will be generated without also giving
a similar operator with the 10 appearing next to the $16_3$.

Having displayed the flavor symmetry structure which leads to the desired
operators we now discuss corrections to the mass matrices. The first type of
corrections comes from
other operators generated via diagrams similar to those in figure 21: for
example the operator $O_{13} = 16_1 \; 45_1^3 \; 10 \; 45_1 \; 16_3$.
Whether this operator makes an important contribution depends on the relative
size of the ratios $v_{10}/M_P$ and $v_5/v_{10}$. Another operator which is
generated from integrating out the superheavy $\bar{16}_{H_2}$ is
the $(16_2^\dagger \; S^\dagger 45_{B-L} \; 16_3)_D$. At first sight the
wavefunction mixing generated by this operator is dangerous. However the new
terms induced are actually higher order, as shown by the exact diagonalization
performed below.
Clearly the family symmetries may give fermion mass matrices which are
dominated by just four operators, but corrections from other operators are
generically expected.

A second type of correction to the light fermion mass matrices arises even in
the limit that all operators beyond the four dominant ones are neglected.
We are considering theories with both superheavy and light 16s.
The diagrams of the form of those in figure 21 represent an
approximation to the process of diagonalizing the mass matrices.
We illustrate this in the above example
by performing an exact diagonalization
of the mass matrices of the heavier two families.
In this sector (including all the relevant heavy
states) there are 2 ${\bar {16}}$'s and 4 16's which we denote as follows:
 $$ {\bar f}_i = {\bar {16}}_{H_i} ,  \ i= 1,2 \eqno(A1.2)$$
$ f_a, \ a= 1,...,4 $ where
$$ f_1 = 16_{H_1}, \ f_2 = 16_{H_2}, \ f_3 = 16_2, \ f_4 = 16_3 \eqno(A1.3)$$
and $f_3,f_4$ are to leading order the light second and third families,
respectively.  The fundamental mass term in the Lagrangian is given by
$$  {\bar f}_i \ m_{ia} \ f_a   \eqno(A1.4)$$
with the $ 2 \times 4 $  mass matrix
$$ m_{ia} = \pmatrix {45_1 & 45_{B-L} & 0 & 0 \cr
                       0   &  45_1  &  S  &   45_{B-L} } \eqno(A1.5) $$

The quadratic mass matrix $m^2 \equiv  \Sigma_{i=1}^2 \ m_{ia} \ m^*_{ib}$
is hermitian and can be diagonalized by a unitary transformation.  In this
case we have
$$ m^2 = \pmatrix{ 45_1^2 & 45_1 \ 45_{B-L} & 0 & 0 \cr
                   45_1 \ 45_{B-L} & 45_1^2 + 45_{B-L}^2 & 45_1 \ S & 45_1
\  45_{B-L}\cr
  0 & 45_1 \ S &  S^2 &  S \  45_{B-L} \cr
0 & 45_1 \  45_{B-L}  & S \  45_{B-L} &  45_{B-L}^2}  \eqno(A1.6) $$
where for simplicity the VEVs are taken to be real.  The eigenstates are
given by
$$  f'_a = V_{ab} \ f_b ~{\rm and}~  f_a = V^*_{ba} \ f'_b \eqno(A1.7)$$
If we denote the massless eigenstates by $f'_3$ and $f'_4$, we find
the exact solutions

$$ f_4 = V^*_{44} \ f'_4 + V^*_{34} \ f'_3 + ...   \eqno(A1.8a)$$

$$ f_3 = V^*_{43} \ f'_4 + V^*_{33} \ f'_3 + ...   \eqno(A1.8b)$$

$$ f_1 = V^*_{41} \ f'_4 + V^*_{31} \ f'_3 + ...   \eqno(A1.8c)$$
with the ellipsis representing the contribution of the massive modes and

\begin{eqnarray}
V_{44} =  {1 \over \sqrt{1 + \alpha_1^2 + \alpha_2^2 + \alpha_3^2}}
& V_{34} = 0 \nonumber\\
  V_{43} = {\alpha_3 \over \sqrt{1 + \alpha_1^2 + \alpha_2^2 + \alpha_3^2}}
& V_{33} = { 1 \over \sqrt{1 +\beta_1^2 + \beta_2^2}} \nonumber \\
V_{41} = {\alpha_1 \over \sqrt{1 + \alpha_1^2 + \alpha_2^2 + \alpha_3^2}}
& V_{31} = {\beta_1 \over \sqrt{1 +\beta_1^2 + \beta_2^2}} \nonumber
\end{eqnarray}
where
\begin{eqnarray}
 \beta_1 = {45_{B-L} \ S \over 45_1^2} & \beta_2 = -{S \over 45_1}
\nonumber \\
\alpha_1 = -{45_{B-L} \over 45_1} \ \alpha_2 & \alpha_2 = - { 45_{B-L}
\over 45_1} \
\left( 1 + {S^2 \over 45_1^2} + ({45_{B-L} \ S \over 45_1^2})^2
\right)^{-1} \nonumber \\
\alpha_3 = - ( \beta_1 \ \alpha_1 + \beta_2 \ \alpha_2) \nonumber
\end{eqnarray}

We can now evaluate the electroweak symmetry breaking mass terms due to the
Higgs scalars sitting in the 10 dimensional representation.  We have

$$  16_3 \ 10 \ 16_3 + 16_2 \ 10  \ 16_{H_1}  = f_4 \ 10 \ f_4 + f_3 \ 10 \
f_1 \eqno(A1.9)$$
We now use equation (A1.8) to rewrite (A1.9) in terms of the massless fields
$f'_4, f'_3$. This gives us our effective operators $O_{ij}$. We also relabel
the states $f'_4 \rightarrow 16_3$ and $f'_3 \rightarrow 16_2$ in order to
obtain formulae similar to that of equation (A1.1).  We have

$$ O_{33} = 16_3 \ \left( V^*_{44}\ 10 \ V^*_{44}  +  V^*_{43} \ 10 \
V^*_{41} \right) \ 16_3   \eqno(A1.10a) $$

$$ O_{23} = 16_2 \ \left( V^*_{31} \ 10 \ V^*_{43} + V^*_{33} \ 10 \ V^*_{41}
\right) \ 16_3  \eqno(A1.10b) $$

$$ O_{22} = 16_2 \ V^*_{33} \ 10 \ V^*_{31} \ 16_2 \eqno(A1.10c) $$
To leading order we obtain the operators, $O_{33} ({\rm eqn}(2.1)), O_{22}
({\rm eqn}(5.6f))$ and $ O_{23}$, the third operator with $\chi =8/9$ in
$({\rm eqn}(5.12c).$  The relative corrections to the leading order results
are of order ${ 45_{B-L}^2 \over 45_1^2}$ or ${S^2 \over 45_1^2}$ which may be
as large as 10 \%.  It should also be noted that these corrections include
Clebsches which differ for up and down quarks and leptons.  Finally it is
instructive to notice that if the ``22" element vanishes in the up mass matrix
to leading order then this result is true for the exact mass matrix as
obtained above (eqn(A1.10c)).

 \vskip 9pt
\noindent{\bf Appendix 2}
\vskip 9pt

In this appendix we use experimental data to exclude
theories with the heavy 2 $\times$ 2 sector
given only in terms of two operators,  $O_{33} + O_{23}$, where  $O_{33}$ is
given in (2.1) and  $O_{23}$ has the form of (2.2) and has
dimension $\le 6$.
The operator  $O_{33}$ works very well for the heaviest generation. The
difficulty arises in finding a suitable  $O_{23}$ to account for the four
observables $V_{cb}^2, m_c/m_t, m_s/m_b$ and $m_\mu/ m_\tau$. This is largely
because the first two of these quantities are considerably smaller than the
second two, as shown in (5.4e).

The Yukawa matrices of the two heavy families $16_3$ and $16_2$ at the GUT
scale are given by:
$$
{\bf U} = \pmatrix{ 0&x'_u B\cr x_u B & A}
{\bf D} = \pmatrix{0&x'_d B\cr x_d B&A}
$$
$$
{\bf E} = \pmatrix{ 0 &x'_e B\cr x_e B&A}
\eqno(A2.1)
 $$
where the $x_i$ and $x_i'$ are the Clebsches resulting from
the operator $O_{23}$. The two operator coefficients, A and B, have been made
real by phase rotations on $16_2$ and $16_3$.
{}From these Yukawa matrices it follows that at $M_G$:

$${\lambda_t \over \lambda_c} x_u x'_u = {\lambda_b \over \lambda_s} x_d x'_d
= {\lambda_{\tau} \over \lambda_{\mu}} x_e x'_e = {|x_u - x_d|^2 \over
V_{cb}^2} \equiv  {A^2 \over B^2} \eqno(A2.2) $$
giving the GUT relations
$$ \xi \equiv {x_d x'_d \over x_u x'_u} = {\lambda_t \over \lambda_b}
{\lambda_s \over \lambda_c}
\eqno(A2.3)$$
$$\chi \equiv { |x_u - x_d| \over \sqrt{|x_u x_u'|}} =
 V_{cb} \sqrt{{\lambda_t \over \lambda_c}} \eqno(A2.4)$$
Using the analysis of section 4 to scale the quantities on the right-hand sides
to low energies gives
$$ \xi  = {m_t \over m_c}{m_s \over m_b} { \eta_b \eta_c \over \eta_s}
\; e^{2 (I_t - I_b)} \eqno(A2.5)$$

$$\chi= V_{cb} \sqrt{{m_t \over m_c}}  \sqrt{\eta_c}
\; e^{{1 \over 2}(I_t - I_b)} \eqno(A2.6)$$
Using experimental values for $m_c, m_b$ and $V_{cb}$, together with the
prediction for $m_t(\alpha_s)$ which follows from $O_{33}$ and the values of
the scaling factors $\eta_i(\alpha_s), I_t(\alpha_s)$ and $I_b(\alpha_s)$ we
find the numerical values:
$$
0.55 < \chi < 0.92
$$
$$
4 { m_s \over 100 MeV} < \xi <7 { m_s \over 100 MeV} \eqno(A2.7)
$$
where variation is due to $0.11 < \alpha_s < 0.13$.
For the purposes of this appendix we will discard models only if they are in
gross disagreement with (A2.7), in particular if they are outside the range
$$
0 < \chi < 2
$$
$$
1 < \xi < \infty \eqno(A2.8)
$$
{}From the table it is
immediate to see that the only dimension 5 operators giving $\xi > 1$  are
$ 16_2 {45_1 \over M} 10 \; 16_3 $ or $16_2 10 {45_1 \over
M} 16_3$: they both give $\xi = 3$. However the former gives $\chi = 0$ and the
latter $\chi = 4$ so they are not acceptable.

Now consider dimension 6 operators. Note that  \ul{45}$_{T_3R}$ may not appear
since it gives $m_c = m_s = m_\mu = 0$. Furthermore the appearance of
\ul{45}$_{B-L}$ is irrelevant: it does not distinguish between the up and down
sectors and hence will not affect $\xi$ or $\chi$. Hence we need only consider
\ul{45}$_1$ and  \ul{45}$_{24}$. Since  \ul{45}$_{24}$ leads to $\xi < 1$,
there
must be more  \ul{45}$_1$s appearing in the numerator than in the denominator
to ensure $ \xi
> 1$. At dimension 6 this implies that either one or two  \ul{45}$_1$s are in
the numerator and none in the denominator, which already implies $\chi \geq 1$.
In fact, dimension 6 models constructed from  \ul{45}$_1$ and  \ul{45}$_{24}$
all have $\chi = 0$ or $\chi \geq 2$ and hence are excluded.

These arguments show that a correct description of the heavier two families
requires the addition of a third operator. The alternatives are:

A) add an operator mixing  $16_2$ with $16_2$  (`` 22 " models);

B)  add one more operator mixing $16_2$ with $16_3$  (`` 23 " models); or

C) add one more higher dimension operator mixing $16_3$ with $16_3$.

It is easy to see that the last option (C) will not help.  The higher
dimension operator will make a $\le 20 \%$ change in the $16_3 16_3$ entries
of the mass matrices.  Such a change will alter the allowed ranges of $\xi$ and
$\chi$ given in (A2.7) by $\approx 20 \%$. However we have already proven that
no model met the very much weaker constraints of (A2.8), and hence (C) will not
yield an acceptable model.

This concludes the proof that we need to consider theories of type A or B.

What about anti-SU(5) theories?  As we shall see later they are excluded by
considering the ``12" entry of the mass matrices.

 \vskip 9pt
\noindent{\bf Appendix 3}
\vskip 9pt

In this appendix we show that for the theories introduced in section 3,
which have fermion masses described by four $SO(10)$ operators, the operator
which gives rise to the 12 element of the mass matrices is unique.
This relies only on the 33 operator being $16_3\ 10\ 16_3$, and
does not depend on the form of the 22 or 23 operators.
In fact it applies to both the ``22'' and ``23'' textures.

{}From the determinant of the light $2\times 2$ sector of the mass matrices
one derives
$$
{m_dm_s\over m_em_\mu} = \left({m_b\over m_\tau}\right)^2 {z_dz_d'\over
z_ez_e'}\ {\eta_d\eta_s\over \eta_b^2}\ {\eta^2_\tau\over \eta_e\eta_\mu}\:
 e^{2I_t +6I_b - 6I_\tau}\eqno(A3.1)
$$
Inputing $m_e, m_\mu, m_\tau, m_b$ and $m_s/m_d=20$ leads to a result for the
strange mass:
$$
m_s \simeq 180 MeV \sqrt{{z_dz_d'\over z_ez_e'}}{ \sqrt{\eta_d\eta_s}\over
\eta_b}\: { e^{I_t+3I_b-3I_\tau} \over 2.3}.\eqno(A3.2)
$$
Thus a strange mass of 180 MeV results only if ${z_dz_d'\over z_ez_e'}$
is close to unity (in fact 1 to 1.7 as $\alpha_s(M_Z)$ decreases from 0.13
to 0.11).
This is not a surprising result.
Once $(\lambda_\mu/\lambda_s)_G =3$ has been arranged, we know from Georgi
and Jarlskog's work that the 12 entries should be the same for the down and
lepton matrices. Given that several \ul{45} VEVs are expected to occur in
the 12 operator, how can we arrange for $z_dz_d' =z_ez_e'$?
One possibility is that the 12 operator involves only \ul{45}$_1$.
In this case the Clebsches $z_i, z_i'$ do not feel $SU(5)$ breaking
in the e and d sectors so
that $z_dz_d' = z_ez_e'$ is automatic.
At first thought one might guess that there would be other ways to arrange a
product of many \ul{45}s (involving \ul{45}$_{B-L}$ \ul{45}$_{T_3R}$ and
\ul{45}$_{24}$ as well as \ul{45}$_1$) such that $z_dz_d' \simeq z_ez_e'$
occurs as an accident.
In fact it is very straightforward to see that this is impossible.
{}From the Clebsch table we see that the appearance of \ul{45}$_{T_3R}$ in the
12 operator is forbidden as it leads to $z_iz_i'=0$.
Furthermore the appearance of either a \ul{45}$_{24}$ or a \ul{45}$_{B-L}$
leads to a factor of 9 appearing in $z_ez_e'/z_dz_d'$.
Since there is no possibility of any non-zero factor less than unity
contributing to $z_ez'_e/z_dz'_d$, we can immediately use the result (A3.2)
to conclude that the 12 operator can involve only \ul{45}'s with VEVs
pointing in the $SU(5)$ preserving direction: \ul{45}$_1$ ie:
$$
O_{12} = 16_1 \left( {45_1\over M}\right)^n 10
\left( {45_1\over M}\right)^m
16_2\eqno(A3.3)
$$
where $n, m$ are positive or negative integers.

We can also use the strange mass prediction of (A3.2) to exclude any model
where $SO(10)$ is broken to the anti-$SU(5)$ subgroup, so \ul{45}$_1$ lies
in an anti-$SU(5)$ singlet direction.
In this case the $z_i, z'_i$ must be replaced by the $\overline{z}_i,
\overline{z}'_i$ relevant to the anti-$SU(5)$ subgroup.
Thus $m^2_s$ is proportional to the
$\overline{z}_d\overline{z}_d'/\overline{z}_e\overline{z}_e' = z_u z_u'
/z_\nu z_\nu'$.
{}From the Clebsch table one immediately sees that any \ul{45} VEV gives this
ratio very far from unity: either zero or $\geq 9$.
The remaining possibility, $n=m=0$, we exclude because it is completely
unreasonable that in our framework the masses of the lightest generation
would arise from a dimension 4 renormalizable operator.

A stringent restriction on $n,m$ can be obtained by considering the
determinant of the light 2 $\times 2$ submatrices in the up and down sector:
$$
{m_um_c\over m_dm_s} = \left({m_t\over m_b}\right)^2 {z_uz'_u\over z_dz_d'}
{\eta_u\eta_c\over \eta_d\eta_s} \eta^2_b\: e^{ 4(I_t - I_b)}\eqno(A3.4)
$$
{}From the Clebsch table
$$
{z_uz_u'\over z_dz'_d} = \left({1\over 3}\right)^{n+m}\eqno(A3.5)
$$
Hence we find
$$
{m_u\over m_d} \simeq .9 \left({1\over 3}\right)^{n+m-6} \left({m_s\over 180
MeV}\right)\eta \eqno(A3.6)
$$
where $\eta = (.6, 1, 1.1)$ for $\alpha_s(M_Z)=(.11, .12, .13)$.
Thus even allowing the very wide range 0.2 $< {m_u\over m_d} < 1.5$, one can
conclude that
$$
n+m = 6 \ \hbox{or} \ 7\eqno(A3.7)
$$

The third and final part of the argument follows from the prediction for the
Cabibbo angle.
{}From the KM matrix of eq. 4.3 one finds $\sin \theta_c=|s_1+s_2e^{-i\phi}|$.
The 12 and 21 entries of the up mass matrix are symmetric so that $s_2 =
\sqrt{\lambda_u\over \lambda_c} = \sqrt{ {\eta_c\over \eta_u}{m_u\over
m_c}}$.
However, this symmetry is not necessarily true for the down matrix so that
$s_1= \sqrt{ \left({1\over 3}\right)^{n-m} {m_d\over m_s}}$.
Because $s_2$ is considerably less than sin $\theta_c$ one must have $s_1$
in the neighborhood of sin $\theta_c$.
As $\sqrt{m_d/m_s} = .22 \pm .02$ is actually centered on sin $\theta_c$,
having $n \neq m$ makes $s_1$ too different from sin $\theta_c$ to obtain a
successful result.
In fact the only case with $n\neq m$ which comes even close to working is
$n=4$ and $m=3$.
This case is excluded by combining the Cabibbo angle constraint with
the result of (A3.6) which gives $m_u/m_d < 0.4$.
Hence we are left with the case $n=m=3$ and the unique operator
$$
O_{12} = \psi_1 \left({45_1\over M}\right)^3 10 \left({45_1\over M}\right)^3
\psi_2\eqno(A3.8)
$$

\vskip 9pt
\noindent{\bf Appendix 4 Supersymmetric threshold corrections to Yukawa
couplings.}
\vskip 9pt

The top mass prediction of Figure 1 is large: about two standard deviations
above the central value extracted from precision electroweak data in the MSSM.
If searches for the top quark prove
that the top is indeed this heavy, the case for the scheme proposed
in this paper will be strengthened, but by no means proved. If the top quark is
found to be much lighter, say 140 GeV, it is important to know to what extent
the framework of this paper is destroyed.

There are several mechanisms which could affect, to varying degrees, our top
mass prediction.
\begin{itemize}

\item Even if none of the assumptions listed in section 3 are relaxed, there is
an effect which can perturb the top prediction. The equality $\lambda_t =
\lambda_b = \lambda_\tau$ is exact at the GUT scale
in the limit that the higher dimension
operators are ignored. However the operator $O_{23}$ perturbs this relation, as
shown in eq. (6.1). In the ``22" models this is negligable (except for model
9), but in the ``23" models it is important, as can be seen in the predictions
for the top mass in Figures 12, 13 and 14.

Other mechanisms affecting $m_t$ require that at least one of the assumptions
of section 3 be relaxed.

\item It could be that GUT scale threshold effects, such as
an additional non-renormalizable operator, upset the GUT
scale relation  $\lambda_t = \lambda_b = \lambda_\tau$. Such perturbations
could
arise when assumptions 3, 6 or 7  of section 3 are
relaxed, and are considered in appendix 5.

\item The supersymmetric threshold corrections to  $\lambda_b$ could be
significant if the last assumption of section 3
is relaxed, and it is this possibility that we explore in this appendix.

\end{itemize}

While the tree-level contribution to the b and $\tau$
masses are proportional to the
small VEV $v_1$, there are one loop
diagrams with internal superpartners which are
proportional to $v_2$. Hence for $\tan \beta \approx 50$ the radiative
corrections are naively $\approx 50\%$ rather than $\approx 1\%$.
Such radiative corrections imply that the quark and lepton mass predictions
could depend on the soft supersymmetry breaking parameters which appear in
these loop diagrams. This would make it impossible to give precise numerical
predictions until the superpartner sector has been discovered and studied.

As an example, consider a radiative correction to the b quark mass which is
35\% of the
tree result, and which subtracts from the tree result. This means that the tree
b quark Yukawa coupling, which is the input needed for the top mass prediction,
is now 35\% larger, resulting in a decrease in the top mass
predicted from operator (2.1) by about 30 GeV.
The leading radiative corrections to the b quark mass are from gluino and
Higgsino exchange diagrams \cite{HRS} shown in figure 22.
In the gluino exchange diagram there is
a trilinear scalar interaction proportional to $\mu$ while in the Higgsino
exchange diagram it is proportional to $A$. For small $\mu$ the gluino exchange
diagram yields
$$
{\delta m_b \over m_b} = { 2 \alpha_s \over 3 \pi} \tan \beta { \tilde{m} \mu
\over m^2} I({\tilde{m}^2 \over m^2}) \eqno(A4.1)
$$
where, in the limit that the two b squarks are nearly
degenerate with mass m, and the gluino has mass $\tilde{m}$
$$
I(x) = { 1 + {x \over 1 - x} \ln x \over 1 - x} \eqno(A4.2)
$$

We can divide the parameter space of the MSSM into two regions according to
whether these one-loop radiative corrections cause substantial changes to the
tree-level predictions. If the corrections are substantial then some of the
fermion mass predictions will depend on the parameters $m, \tilde{m}, \mu$ and
$A$. The region where these corrections can be ignored is, roughly speaking,
that where $\mu$, and
probably $\tilde{m}$ and $A$, are less than the squark mass m.
This region of parameter space is preferred when $\tan \beta$ is large
\cite{HRS}.
The Higgs potential of the MSSM involves three parameters: $\mu_1^2,
\mu_2^2$ and $\mu_3^2$. One combination of these is determined by the
constraint of the observed Z mass. The other two parameters can be
taken as $\tan \beta$ and the pseudoscalar mass $m_A$ \cite{KZ}. For large
$\tan \beta$ one finds: $\mu_3^2 = m_A^2 / \tan \beta$. Thus large $\tan \beta$
forces $\mu_3^2 = \mu B$ to be small, which is most naturally accomplished by
a combination of approximate Peccei-Quinn ( $\mu/m \approx \epsilon_{PQ} \ll
1$) and R ($ \tilde{m}/m \approx A/m \approx B/m \approx \epsilon_R \ll 1$)
symmetries. The potentially large radiative corrections to $m_b$, such as in
equation (A4.1), are all proportional to $\epsilon_{PQ} \epsilon_R \tan \beta$
which is expected to be of order unity, because it is the above approximate
symmetries which are the origin of small $\cot \beta$: $\cot \beta \approx
\epsilon_{PQ} \epsilon_R$. Hence typically these radiative corrections are not
expected to be sufficiently large to substantially alter the predictions of the
flavor parameters and are ignored in the body of this paper.
Note however, that we
cannot {\em exclude} the case where they are significant, which would happen
for example if $\tilde{m} / B \approx 5$. In this case the top quark could be
much lighter than our prediction shown in Figure 1. Nevertheless, the other
predictions survive with slight modifications, as discussed  later in
this Appendix.

Diagrams similar to those which correct the b quark mass
(one of the external b quarks
becomes an s quark, and a photon is attached) lead to an amplitude for b
$\rightarrow s \gamma$ linear in $\tan \beta$. Thus the possibility of sizable
radiative corrections to $m_b$ can be probed via $b \rightarrow s \gamma $
\cite{HRRS}.

The suggestion \cite{HRS} that the small value of $m_b/m_t$ is due to
approximate Peccei-Quinn and R symmetries deserves further attention, both from
the viewpoint of the radiative mechanism for electroweak symmetry breaking and
the resulting experimental signatures (both charginos are expected to have
light masses $\approx 100$ GeV). What are the origins of these approximate
symmetries? An approximate Peccei-Quinn symmetry requires $\mu$ to be less
than the
size of the supersymmetry breaking parameters. These scales are logically
independent; their equality is often called the ``$\mu$'' problem. In one
solution of this problem the $\mu$ parameter arises as a radiative correction
from the supersymmetry breaking scale, and is expected to be small
\cite{H84}. The approximate R symmetry, on the other hand, must result from the
particular pattern of supersymmetry breaking.

Now let us consider radiative corrections to our results in more detail.
The diagrams of figure 22 lead to supersymmetric threshold corrections to
$\delta m_b/m_b$, which are naively $\alpha_s / 4 \pi$ and $\lambda_t^2 / 16
\pi^2$ respectively, which are both $\approx 1 \%$. These diagrams yield
\cite{HRS}:
$$
{\delta m_b \over m_b} \sim {2 \alpha_3 \over 4 \pi}
\left( {m_{1/2} \over B} \right)
\left( {m_A^2 \over m_o^2} \right) \eqno(A4.3a)
$$
$$
{\delta m_b \over m_b} \sim { \lambda_t^2 \over 16 \pi^2}
\left( {A_t \over B} \right) \left( {m_A^2 \over m_o^2} \right) \eqno(A4.4b)
$$
where $A_t, B$ and $m_{1/2}$ are supersymmetry breaking parameters, $m_A$ is
the mass of the pseudoscalar of the MSSM and $m_o$ characterizes the
mass scale of the degenerate scalar superpartners.
The most natural expectation is that  $A_t \sim B \sim m_{1/2}$ and $m_A <
m_o$ in which case ${\delta m_b \over m_b} < 2$ \%, 2/3 \% respectively, so
that
the top mass is changed by less than 3 GeV. The most natural expectation is
that these radiative corrections do not substantially change the predictions
shown in figure 1. Nevertheless, it is not
extremely unnatural for, say, $m_{1/2} /B  \sim$ 5 or 10, which could lower the
top mass prediction by 15 to 30 GeV. The sign of the diagrams is unknown
because the sign of $\mu$ is unknown, in this section we are interested in the
sign which leads to a decrease in $m_t$.

The effects of $\delta m_b$ can be understood as follows: the b Yukawa coupling
gets multiplied by a factor $(1-\xi)$, so that in eq. (4.10) $\eta_b
\rightarrow \eta_b (1-\xi)$. Thus the changes are identical to those that would
occur if the physical b mass, $m_b$, were multiplied by $1/(1-\xi)$. For $\xi$
positive this increases $m_b$, and the resulting decreases in $M_t$ and $\tan
\beta$ can now be read from figures (1a) and (1b).

The leading effects of this $\lambda_b$ threshold correction
on the other predictions of the ``22" models can now be
obtained from eqs. (5.33) - (5.37) by multiplying $m_b$ by  $1/(1-\xi)
\sim (1+\xi)$ and using the corrected value for $m_t$.
Subleading corrections arise because the Yukawa coupling A has changed,
inducing a change in the integrals $I_{t,b,\tau}$.

However, some of these predictions are also affected by threshold corrections
to other elements of the down Yukawa matrix which result from figures 22a and
22b. In this appendix we consider these corrections in the following
approximation. We ignore the RG scaling of the down type squark mass matrix,
and of the down type A matrix,
except for the effect which makes $\tilde{b}$ not degenerate with $\tilde{s}$
and $\tilde{d}$. In this approximation the gluino exchange diagram conserves
flavor, so that its only additional effect is to give equal threshold
corrections to $\lambda_s$ and $\lambda_d$. This can be accounted for by
multiplying the predictions for $m_s$ and $m_d$ by $1-\xi'$, where $\xi'$ has
the same sign as $\xi$ and is equal to it in the limit of squark degeneracy.
Hence the leading effects of the gluino exchange diagram are

1) $m_t$ multiplied by $(1-.8\xi)$.

2) $\tan \beta$ multiplied by $(1-1.7\xi)$.

3) $V_{cb}$ multiplied by $(1+.4\xi)$.

4) $m_s$ multiplied by $(1+\xi)/(1+\xi')$.

5) $m_s/m_d$ unchanged.

6) $m_u/m_d$ multiplied by $(1-2.6\xi)/(1+\xi')$

7) $V_{ub}/V_{cb}$ multiplied by $(1-.8\xi)$.

8) J unchanged.

We see that if the supersymmetric threshold corrections are surprisingly big,
for example $\xi = $ 0.1 to 0.2,
our top mass prediction is largely lost, but the other predictions are simply
perturbed. The most interesting change is for $m_u/m_d$, since it is decreased
and this improves the agreement with experiment in all ``22" models.

The higgsino exchange diagram of figure 22b is typically smaller than the
gluino exchange diagram of figure 22a. The flavor structure of this diagram is
more complicated than for the gluino exchange diagram because the charged
Higgsino vertices contain a factor of the CKM matrix. Nevertheless, the diagram
is extremely small unless the  exchanged scalar is from the third generation,
and the only operators generated
significantly are: $b b^c, s b^c$ and $ d b^c$. The
first contributes to $\xi$ as above, while the others perturb $V_{cb}$ and
$V_{ub}$.

\vskip 9pt
\noindent{\bf Appendix 5 Threshold effects from the grand unification scale.}
\vskip 9pt

In this appendix we consider the effects of GUT scale threshold corrections
perturbing the boundary conditions on the Yukawa matrices. This could occur in
many ways; we will just consider some simple cases for which there may be some
motivation.

The Yukawa couplings of the
third generation, $U_{33}, D_{33}$ and $E_{33}$ can be perturbed by the
presence
of an additional non-renormalizable operator. This could allow $\lambda_t >
\lambda_b$ at the GUT scale, which may be the easiest way to allow radiative
electroweak symmetry breaking. If this extra operator gives contributions to
the Yukawa couplings which preserve $SU(5)$, for example
$$
O_{33}' = 16_3 \; 10 \; 45_1 \; 16_3 \eqno(A5.1)
$$
then the t Yukawa can increase slightly, by a correction $v_{10}/ M_P$,
compared
with the b and $\tau$ couplings, which remain equal. Such a perturbation in the
boundary conditions leads to a very small change in the top mass prediction
\cite{HRS}: about 2 GeV for  $v_{10}/ M_P = 0.2$. Other predictions are also
affected insignificantly. If $45_1$ is replaced by $45_Y$ in eq. (A5.1), the b
and $\tau$ couplings will no longer be equal at the GUT scale. The relation
(4.10) for $m_b/m_\tau$ will be multiplied by some correction factor,
$(1-\xi)$,
and this can substantially decrease the
top mass and perturb other predictions as
described in Appendix 4.

Finally, we consider the violation of assumption 3, that both Higgs doublets of
the MSSM lie entirely in a single 10. Suppose that all Yukawa interactions are
generated by the couplings of a single 10, but this 10 contains only components
of the two light Higgs doublets. The other components must come from some other
representation $10', 120, 126,$ etc. The effect is to multiply the Yukawa
matrix
{\bf U} by one mixing angle and {\bf D} and {\bf E} by another. The mixing
angle which multiplies {\bf U} can be absorbed into the coupling constants, so
the net result is to simply multiply {\bf D} and {\bf E} by some number
$\zeta$. The effect of $\zeta$ is to rescale $\tan \beta$. For example eq.
(4.10) for $m_b/m_\tau$ is unchanged, but in eq. (4.11) $\cos \beta$ must be
multiplied by $\zeta$. With $\zeta$ very small there is no need for $\tan
\beta$ to be large, so electroweak symmetry breaking could occur without any
fine tuning. The one extra free parameter, $\zeta$, means that $\tan \beta$ can
no longer be predicted. Even though the top Yukawa can be predicted, the
top quark mass prediction, which is proportional to $\lambda_t \sin \beta$,
is lost. The remaining 6 flavor predictions for each of the ``22" models
are given by the same expressions as before. The numerical values for the
predictions will change slightly due to the change in
the integrals $I_{t,b,\tau}$.

\newpage
\begin{description}
\item[\it Figure 1a:] The physical top quark mass prediction is plotted
as a function of the $\bar{MS}$ value of the bottom quark mass for
$\alpha_s(M_Z) = 0.110 -0.126$.  $M_t$ is an increasing function of $\alpha_s$.
The circles (diamonds) represent points where the GUT scale Yukawa coupling
is equal to 2 (3).
\item[\it Figure 1b:] The ratio of vacuum expectation values,
$\tan\beta$, is plotted as a function of the $\bar{MS}$ value of the
bottom quark mass for $\alpha_s(M_Z) = 0.110 -0.126$.  $\tan\beta$ is an
increasing function of $\alpha_s$.  The circles (diamonds) represent
points where the GUT scale Yukawa coupling is equal to 2 (3).

\item[\it Figure 2:] The integrals $I_t$, $I_b$, and $I_\tau$.
$I_t$ (solid curve), $I_b$ (dashed curve), and $I_\tau$ (dotted curve),
are plotted as of function of the $\bar{MS}$ value of the bottom quark mass.

\item[\it Figure 3:] The renormalization group mass enhancement factors
$\eta_i$ are plotted as a function of $\alpha_s(M_Z)$.
$\eta_i = m_i(m_i)/m_i(m_S)$ for $i = c,b$ and
$\eta_i = m_i(1 GeV)/m_i(m_S)$ for $i = u,d,s$

\item[\it Figure 4:] $V_{cb} = \chi \sqrt{m_c/m_t}$ plotted as a function
of $\alpha_s(M_Z)$ for $\chi = 2/3,5/6,8/9$ and 1.  The region between
the dashed (dotted) curves is the range of predictions for $V_{cb}$
for $m_c(m_c) = 1.27 \pm 0.05 (0.1)$ GeV and
$m_b(m_b) = 4.25 \pm 0.1 (0.2)$ GeV.  The solid curve gives the prediction
for $V_{cb}$ with central values of $m_c$ and $m_b$.  Some of the curves
are not continuous across the figure because the GUT scale Yukawa coupling
becomes nonperturbative for small $m_b(m_b)$ and large $\alpha_s$ (See
figure 1a).

\item[\it Figure 5a:] Light quark mass ratios for the $\chi = 2/3$ models.
The mass ratio $m_s/m_d$ is plotted as a fuction of $m_u/m_d$.  These values
are obtained for the range of inputs: $m_b(m_b)= 4.25\pm 0.1$, $m_c(m_c) =
1.27\pm0.05$ and $\alpha_s(M_Z) = 0.110-0.125$. The dependence of $m_u/m_d$ on
these input parameters can be obtained from equation 5.35, or, more accurately,
from Figures 6 and 7 for models 9 and 6 respectively.
The quarter ellipse is the second order chiral perturbation theory prediction
for the light quark mass ratios of Eq 6.2.
The dashed (dotted) ellipses are obtained
when the $Q$ in Eq 6.2 is assumed to be uncertain by 13\% (26\%).

\item[\it Figure 5b:] Light quark mass ratios for the $\chi = 5/6$ models.
The mass ratio $m_s/m_d$ is plotted as a fuction of $m_u/m_d$.  These values
are obtained for the range of inputs: $m_b(m_b)= 4.25\pm .1$, $m_c(m_c) =
1.27\pm.05$ and $\alpha_s(M_Z) = .110-125$. The dependence of $m_u/m_d$ on
these input parameters can be obtained from equation 5.35, or, more accurately,
from Figures 6 and 7 for models 9 and 6 respectively.
The quarter ellipse is the second order chiral perturbation theory prediction
for the light quark mass ratios of Eq 6.2.
The dashed (dotted) ellipses are obtained
when the $Q$ in Eq 6.2 is assumed to be uncertain by 13\% (26\%).

\item[\it Figure 5c:] Light quark mass ratios for the $\chi = 8/9$ models.
The mass ratio $m_s/m_d$ is plotted as a fuction of $m_u/m_d$.  These values
are obtained for the range of inputs: $m_b(m_b)= 4.25\pm .1$, $m_c(m_c) =
1.27\pm.05$ and $\alpha_s(M_Z) = .110-125$.
The quarter ellipse is the second order chiral perturbation theory prediction
for the light quark mass ratios of Eq 6.2.
The dashed (dotted) ellipses are obtained
when the $Q$ in Eq 6.2 is assumed to be uncertain by 13\% (26\%).

\item[\it Figure 6:] $m_s/m_d$ and $V_{cb}$ are plotted as a function
of $m_u/m_d$ for model 9.  The same set of inputs has been used in
both figures: $m_b(m_b)= 4.25\pm 0.1$, $m_c(m_c) =
1.27\pm0.05$ and $\alpha_s(M_Z) = 0.110,0.115,0.120$.
For each value of  $\alpha_s(M_Z)$ the prediction is given by five solid lines,
each of which corresponds to a definite value of $m_b(m_b)$, which steps by $0.
05$ GeV between lines. Along each line the input value of $m_c(m_c)$ is varied.
The predicted value for $m_u/m_d$ decreases as either $m_b(m_b)$ or $m_c(m_c)$
is increased.

\item[\it Figure 7:] $m_s/m_d$ and $V_{cb}$ are plotted as a function
of $m_u/m_d$ for model 6.  The same set of inputs has been used in
both figures: $m_b(m_b)= 4.25\pm .1$, $m_c(m_c) =
1.27\pm0.05$ and $\alpha_s(M_Z) = 0.115,0.120$.
For each value of  $\alpha_s(M_Z)$ the prediction is given by five solid lines,
each of which corresponds to a definite value of $m_b(m_b)$, which steps by $0.
05$ GeV between lines. Along each line the input value of $m_c(m_c)$ is varied.
The predicted value for $m_u/m_d$ decreases as either $m_b(m_b)$ or $m_c(m_c)$
is increased.

\item[\it Figure 8:] $V_{ub}/V_{cb}$  is plotted as a function of
$\alpha_s(M_Z)$
in model 6.  The error bars at the left of the graph represent several
determinations, together with the uncertainties,
of $V_{ub}/V_{cb}$ from 1992 CLEO data on the endpoint spectrum
of semileptonic B decays \cite{DRELL}.

\item[\it Figure 9:] The strange quark mass predictions is plotted as
a function of $\alpha_s(M_Z)$ for model 6.

\item[\it Figure 10:] The range of $\hat{B_K}$ predictions for models 1-9.
The input quantities for this figure were evaluated at a scale $\mu$ where
$\alpha_s(\mu) = 1$.  With the exception
of the $\chi = 2/3$ models, all these determinations agree well with
the lattice estimates of roughly .6.

\item[\it Figure 11a:]  $\sin{2\beta}$ vs. $\sin{2\alpha}$ for the
$\chi = 2/3$ models.  Predictions for the four $\chi = 2/3$ models
are drawn over three different backgrounds representing
allowed regions.  The area inside the
solid curve shows the allowed region for theories in which
the GUT scale Yukawa coupling matrices have vanishing (13), (31), and
(11) entries,
$m_s/m_d$ and $m_u/m_d$ lie between the dotted curves of figure 5,
and $V_{ub}/V_{cb}>.07$.  If this lower bound on $V_{ub}/V_{cb}$ is
reduced to .05 (.03), the background expands to the dashed (dotted)
curve.
The prediction for each model appears as a line, with larger values of $\sin 2
\beta$ corresponding to larger $\alpha_s(M_Z)$.

\item[\it Figure 11b:]  $\sin{2\beta}$ vs. $\sin{2\alpha}$ for the
$\chi = 5/6$ models.  Predictions for the two $\chi = 5/6$ models
are drawn over three different backgrounds representing
allowed regions.  The area inside the
solid curve shows the allowed region for theories in which
the GUT scale Yukawa coupling matrices have vanishing (13), (31), and
(11) entries,
$m_s/m_d$ and $m_u/m_d$ lie between the dotted curves of figure 5,
and $V_{ub}/V_{cb}>.07$.  If this lower bound on $V_{ub}/V_{cb}$ is
reduced to .05 (.03), the background expands to the dashed (dotted)
curve.
The prediction for each model appears as a line, with larger values of $\sin 2
\beta$ corresponding to larger $\alpha_s(M_Z)$.

\item[\it Figure 11c:]  $\sin{2\beta}$ vs. $\sin{2\alpha}$ for the
$\chi = 8/9$ models.  Predictions for the three $\chi = 8/9$ models
are drawn over three different backgrounds representing
allowed regions.  The area inside the
solid curve shows the allowed region for theories in which
the GUT scale Yukawa coupling matrices have vanishing (13), (31), and
(11) entries,
$m_s/m_d$ and $m_u/m_d$ lie between the dotted curves of figure 5,
and $V_{ub}/V_{cb}>.07$.  If this lower bound on $V_{ub}/V_{cb}$ is
reduced to .05 (.03), the background expands to the dashed (dotted)
curve.
The prediction for each model appears as a line, with larger values of $\sin 2
\beta$ corresponding to larger $\alpha_s(M_Z)$.

\item[\it Figure 12:] Predictions for model A.

The variation of the prediction with $\alpha_3(M_Z)$
is shown for: the top quark pole mass $M_t,
\tan \beta$, $m_s(1 GeV), V_{cb}$ and $V_{ub}/V_{cb}$. The prediction for
$m_s/m_d$ versus $m_u/m_d$ is also shown. Superpartners are taken degenerate at
200 GeV. The lines are for contours of $0.5 \sigma,
1.0 \sigma, 1.5 \sigma$ and
$2.0 \sigma$ for input ranges of $m_b(m_b) = 4.25 \pm 0.25,
m_c(m_c) = 1.27 \pm 0.05$ and $\sin \theta_c = 0.2205 \pm 0.0018$. Thus, within
the innermost contour all three inputs $m_b(m_b), m_c(m_c)$ and $\sin \theta_c$
are within $0.5 \sigma$ of their central values. In some regions the $1.5
\sigma$ and $2 \sigma$ contours are indistinguishable.

\item[\it Figure 13:] Predictions for model B (same as figure 12).

\item[\it Figure 14:] Predictions for model C (same as figure 12).

\item[\it Figure 15:] Prediction for $B_K$ in models A, B and C.

The range corresponds to $2 \sigma$ variations of the inputs (see figure 12);
and does not differ greatly from the $ 1 \sigma$ allowed regions. The
superpartners are taken degenerate with mass 200 GeV.

\item[\it Figure 16:] $V_{cb}$ versus $B_K$ in models A, B and C.

The range corresponds to $2 \sigma$ variations of the inputs (see figure 12);
and does not differ greatly from the $ 1 \sigma$ allowed regions. The
superpartners are taken degenerate with mass 200 GeV.

\item[\it Figure 17:]  $V_{ub}/V_{cb}$ versus $B_K$ in models A, B and C.

The range corresponds to $2 \sigma$ variations of the inputs (see figure 12);
and does not differ greatly from the $ 1 \sigma$ allowed regions. The
superpartners are taken degenerate with mass 200 GeV.

\item[\it Figure 18:]  $V_{ub}/V_{cb}$ versus $V_{cb}$ in models A, B and C.

The range corresponds to $2 \sigma$ variations of the inputs (see figure 12);
and does not differ greatly from the $ 1 \sigma$ allowed regions. The
superpartners are taken degenerate with mass 200 GeV.

\item[\it Figure 19:] $B_{B_d} F_B^2$ versus $\alpha_3(M_Z)$ for models A, B
and C.

The range corresponds to $2 \sigma$ variations of the inputs (see figure 12);
and does not differ greatly from the $ 1 \sigma$ allowed regions. The
superpartners are taken degenerate with mass 200 GeV.

\item[\it Figure 20:] $\sin 2 \beta$ versus $\sin 2 \alpha$ for models A, B, C.

The range corresponds to $2 \sigma$ variations of the inputs (see figure 12);
and does not differ greatly from the $ 1 \sigma$ allowed regions. The
superpartners are taken degenerate with mass 200 GeV.

\item[\it Figure 21:] Diagrams generating the operators of model 9.

A) The Yukawa coupling for the third generation.

B) Generation of $O_{23}$ via exchange of superheavy generations.

C) Generation of $O_{22}$ via exchange of superheavy generations.

D) Generation of $O_{12}$ via exchange of superheavy generations.

The superscripts label the charges of the representations under a new U(1)
symmetry.

\item[\it Figure 22:] Supersymmetric threshold corrections to the Yukawa
couplings of the down-type quarks from: a) Gluino exchange b) Charged Higgsino
exchange.

Tildes denote superpartners and V is the CKM matrix.

\end{description}
\end{document}